\documentclass{article}
\usepackage{a4wide}
\usepackage{amscd}
\usepackage{amssymb}
\usepackage{amsthm}
\usepackage{amsmath}
\usepackage{times}
\usepackage{latexsym}
\usepackage{amsfonts}
\usepackage{graphicx, subfigure, graphics}
\usepackage{color}
\usepackage{comment}
\usepackage{authblk} 

\newtheorem{theorem}{Theorem}
\theoremstyle{plain}

\newtheorem{proposition}[theorem]{Proposition}

\numberwithin{equation}{section}
\numberwithin{theorem}{section}
\newcommand{\R}{\ensuremath{\mathbb{R}}}
\newcommand{\N}{\ensuremath{\mathbb{N}}}
\newcommand{\E}{\ensuremath{\mathbb{E}}}
\def\e{{\mathrm{e}}}
\allowdisplaybreaks

\title{Local sensitivity analysis of heating degree day and cooling degree day temperature derivatives prices}
\author{Sara Ana Solanilla Blanco \thanks{Financial support from the project "Energy Markets: Modelling, Optimization 
and Simulation" (EMMOS) funded by the Norwegian Research Council is gratefully acknowledged.}}
\affil{University of Oslo \\ P.O. Box 1053, Blindern \\ N--0316 Oslo, Norway}

\begin{document}
\maketitle

\begin{abstract}
We study the local sensitivity of heating degree day (HDD) and cooling degree day (CDD) temperature futures and option prices with respect to 
perturbations in the deseasonalized temperature or in one of its derivatives up to a certain order determined by 
the continuous-time autoregressive process modelling the deseasonalized temperature in the HDD and CDD indexes. 
We also consider an empirical case where a CAR process of autoregressive order 3 is fitted to New York temperatures
and we perform a study of the local sensitivity of these financial contracts and a posterior analysis of the results. 
\end{abstract}

\section{Introduction}\label{sec:intro}

Weather related risks can be hedged by trading in weather derivatives.
The Chicago Mercantile Exchange (CME) organizes trade in futures contracts written
in weather indexes in several cities around the world.
We focus on the temperature indexes HDD (heating-degree day) and CDD
(cooling-degree day) which measure the aggregation
of temperature below and above a threshold of $65^{\circ}$F over a time period, respectively.
The daily modelling of temperature 
is an approach that can be used to get the non-arbitrage price of temperature derivatives.
A continuous-time function which consists of a deterministic 
term modelling the seasonal cycle 
of temperatures and a noise term modelling uncertainty is fitted
to historical time series of daily average temperatures (DATs). 
Several empirical studies of temperature 
data, see Härdle and López Cabrera \cite{HLC}, Benth and \v{S}altyt\.{e} Benth \cite{BSB-book} and the references therein, have shown that continuous time autoregressive (CAR) models explain very well the statistical properties of the deseasonalized temperature dynamics.
Although this approach requires a model for the instantaneous temperature, it has the advantage that 
the model can be used for all available contracts on the market on the same location.
In Benth and Solanilla Blanco \cite{BenSol-A3}, HDD and CDD futures prices based on CAR temperature dynamics 
and option on these futures are derived theoretically. An approximative model for the HDD and CDD futures is suggested in order
to derive a closed formula for the call option price. The (approximative) formulas for HDD and CDD futures and option prices  
depend on the deseasonalized temperature and its derivatives up to $p-1$, where $p\in\N-\{0\}$
refers to the autoregressive order of the CAR process which models the deseasonalized temperature dynamics of these indexes.

The objective of this paper is to study the local sensitivity of the (approximative) HDD and CDD futures  
and option prices derived in Benth and Solanilla Blanco \cite{BenSol-A3}
with respect to perturbations in the deseasonalized temperature or in one of its
derivatives up to order $p-1$. 
To do so, we consider the partial derivatives of such financial contracts with respect to
these variables evaluated at a fixed point.
Local sensitivity measures parameter importance by considering infinitesimal variations
in a specific variable.

Sensitivity analysis is widely used in mathematical modelling to determine the influence of parameter values 
on response variables. The local sensitivity analysis with partial 
derivatives is a first step to study the response of a model 
to changes in their inputs variables. In mathematical finance there is extensive literature 
of Greeks, which are quantities representing the sensitivity of derivatives prices to a change
in underlying parameters. The sensitivity analysis focused on  
HDD and CDD futures and option prices where the temperature dynamics follows a CAR process has 
not been considered yet. Our contribution is a first analysis of the local sensitivities of such financial contracts
with respect to a perturbation in the deseasonalized temperature or in one of its $p-1$ derivatives.

The paper is structured as follows: in the next section we review results concerning the
arbitrage-free pricing of temperature HDD and CDD futures with measurement over a period and call options written on these. 
We also derive some results to study the sensitivity of these financial contracts to perturbations in the deseasonalized temperature 
or in one of its derivatives up to order $p-1$,
with $p$ being the order of the CAR process used to model the deseasonalized temperature in these indexes. 
In Section 3 we adapt all the results in Section 2 for the case where the measurement time is a day instead. 
In Section 4 we consider a previous empirical study of New York temperatures where the deseasonalized temperature dynamics follows a CAR(3)-process and 
we study local sensitivity of HDD and CDD futures and option prices with a measurement day. 
In Section 5 we include an empirical analysis of the sensitivity of the HDD and CDD futures prices with measurement over a period. 
Finally, in Section 6 we present a conclusion of the results.

\section{Sensitivity of CDD and HDD derivatives prices with measurement period}\label{sec:teo-mp}

In this Section we review results concerning 
pricing futures written on the temperature indexes CDD and HDD defined over a 
measurement period and call options on these temperature futures.
These results are the baseline to develop a study of the sensitivity of these financial contracts 
with respect to changes in the components of the vector function $\mathbf{X}(t)$ involved in 
the definition of the indexes.

Let $(\Omega,\mathcal{F},\{\mathcal{F}_t\}_{t\geq 0},P)$ be a complete probability space, 
the CDD and HDD indexes over a time period $[\tau_1,\tau_2], \tau_1<\tau_2$,
are defined respectively as 
\[
\text{CDD}(\tau_1,\tau_2)=\int_{\tau_1}^{\tau_2}\max(T(t)-c,0)\,dt\,
\]
and 
\[
\text{HDD}(\tau_1,\tau_2)=\int_{\tau_1}^{\tau_2}\max(c-T(t),0)\,dt\,,
\]
where $T(t)$ is the daily average temperature in the location
at time $t$ and the threshold $c$ is $65^{\circ}$F (or $18^{\circ}$C). The temperature is modelled 
as $T(t)=\Lambda(t)+Y(t)$ by means of a seasonal function $\Lambda$
and a CAR($p$)-process $Y(t)=\mathbf{e}_1'\mathbf{X}(t)$ defined as the first component of a multivariate Orstein-Uhlenbeck 
process $\mathbf{X}(t)$ with dynamics 
\begin{equation}
\label{dynamics}
d\mathbf{X}(t)=A\mathbf{X}(t)\,dt+\sigma(t)\mathbf{e}_p\,dB(t).
\end{equation}

We use the notation $\mathbf{z}'$ to denote the transpose
of a vector (or matrix) $\mathbf{z}$. The matrix $A$, which has a particular representation, 
contains the different speeds of mean reversion. We assume the condition of having different eigenvalues with strictly 
negative real part in order to have a stationary model. The function $\sigma(t)$ is the time-dependent volatility of the process.
The arbitrage-free futures price written on a CDD index at time $t\leq\tau_2$ with measurement
period $[\tau_1,\tau_2]$ is defined as 
\begin{equation}
\label{FCDD-def}
F_{\text{CDD}}(t,\tau_1,\tau_2):=\mathbb{E}_Q\left[\text{CDD}(\tau_1,\tau_2)\,|\,\mathcal{F}_t\right],
\end{equation}
where the conditional expectation is taken under
some probability $Q\sim P$. Analogously it can be considered for HDD indexes. 
We choose to work with the probability measure $Q$ considered in Benth and Solanilla Blanco \cite{BenSol-A3} given 
by a Girsanov transform which involves the parameter function $\theta$ referred as the market price of risk and the 
the stochastic process $\sigma$ modeling the volatility.
For a better understanding of this setting we refer Benth and Solanilla Blanco \cite{BenSol-A3},
where it is possible to find also all the results that we present next about futures and call options prices. 
We recall the CDD and HDD futures prices formulas provided
in Proposition.~2.1 and Proposition.~2.3 respectively. For our convenience in this setting we restrict $t\leq\tau_1$, so that 
\begin{equation}
\label{FCDD-formula}
F_{\text{CDD}}(t,\tau_1,\tau_2)=
\int_{\tau_1}^{\tau_2}\Sigma(t,s)\Psi\left(\frac{m_{\theta}(t,s,\mathbf{X}(t))-c}{\Sigma(t,s)}\right)\,ds\,,
\end{equation}
and
\begin{equation}
\label{FHDD-formula}
F_{\text{HDD}}(t,\tau_1,\tau_2)=
\int_{\tau_1}^{\tau_2}\Sigma(t,s)\Psi\left(\frac{c-m_{\theta}(t,s,\mathbf{X}(t))}{\Sigma(t,s)}\right)\,ds.
\end{equation}
Note that $\Psi(x)=x\Phi(x)+\Phi'(x)$, with $\Phi$ being the cumulative standard normal
distribution function and  for $\mathbf{x}\in\R^{p}$
\begin{align*}
m_{\theta}(t,s,\mathbf{x})&=\Lambda(s)+\mathbf{e}_1'e^{A(s-t)}\mathbf{x}+
\int_t^s\mathbf{e}_1'e^{A(s-u)}\mathbf{e}_p\theta(u)\,du \\
\Sigma^2(t,s) &=\int_t^s(\mathbf{e}_1'e^{A(s-u)}\mathbf{e}_p)^2\sigma^2(u)\,du\,.
\end{align*}
If we consider the initial condition $\mathbf{X}(0)=\mathbf{x}\in\R^{p}$, then 
$F_\text{CDD}(t,\tau_1,\tau_2)$ as defined in \eqref{FCDD-def} is a random variable for $t>0$
with all the stochasticity contained in the term $\mathbf{X}(t)$. In our setting   
we loose this condition when for $t\geq0$ we fix $\mathbf{X}(t)=\mathbf{x}\in\R^{p}$ with $\mathbf{x}'=(\texttt{x}_1,\ldots,\texttt{x}_p)$.
In such a case the CDD futures price can be explained as a deterministic function.
Denote by $X_i(t)=\mathbf{e}_i'\mathbf{X}(t)$, for $i=1,\ldots ,p$ the $i$th component of $\mathbf{X}(t)$. 
Note that $Y(t)=\mathbf{e}_1'\mathbf{X}(t)=X_1(t)$. We find that the term $\mathbf{e}_1'e^{A(s-t)}\mathbf{X}(t)$ in the CDD and HDD future prices can be rewritten as follows
\begin{equation}
\label{futures-term-tau1tau2}
\mathbf{e}_1'e^{A(s-t)}\mathbf{X}(t)=\sum_{i=1}^{p}f_{i}(s-t)X_i(t)\,,
\end{equation}
where for $i=1,\ldots,p$
\[
f_{i}(s-t)=\mathbf{e}_1'\exp(A(s-t))\mathbf{e}_i\,,
\]
that is as a linear combination of the components of $\mathbf{X}(t)$.
From now on, we focus on CDD futures prices and call options
written on these. Similar results can be obtained considering the HDD index.
The new notation in \eqref{futures-term-tau1tau2} let us to rewrite the CDD future price formula in \eqref{FCDD-formula} as follows:
\begin{equation}
\label{FCDD-formula-new}
F_{\text{CDD}}(t,\tau_1,\tau_2,\texttt{x}_1,\ldots,\texttt{x}_p)_{\big\vert_{\mathbf{x}=\mathbf{X}(t)}}=\int_{\tau_1}^{\tau_2}\Sigma(t,s)\Psi\left(\frac{m_{\theta}(t,s,\mathbf{X}(t))-c}{\Sigma(t,s)}\right)\,ds\,.
\end{equation}

To answer the question to what extent an infinitesimal change in a component of $\mathbf{X}(t)$ is affecting the behavior of the CDD futures price, we need to consider partial derivatives of this with respect to the components of $\mathbf{X}(t),$ 
say $\texttt{x}_i, i=1,\ldots,p$ to avoid misunderstandings in the notation.
In the next proposition we consider the partial derivatives of the CDD futures price with respect to the components of $\mathbf{x}$.
\begin{proposition}
\label{prop:dFCDD-ttau1tau2}
Let $t\leq \tau_1$, then for $i=1,\ldots,p$ it holds that 
\[
\big(\frac{\partial}{\partial\texttt{x}_i}F_{\text{CDD}}(t,\tau_1,\tau_2,\texttt{x}_1,\ldots,\texttt{x}_p)\big)_{\big\vert_{\mathbf{x}=\mathbf{X}(t)}}=
\int_{\tau_1}^{\tau_2}\Phi\left(\frac{m_{\theta}(t,s,\mathbf{X}(t))-c}{\Sigma(t,s)}\right)\mathbf{e}_1'\exp(A(s-t))\mathbf{e}_i \,ds.
\]
\end{proposition}
\begin{proof}
The proof follows by first exchanging the derivative and the integral and afterwards
applying the chain-rule on the integrand. In this last step consider that $\Psi'(x)=\Phi(x)$
and take into account that $m_{\theta}$ can be rewritten as a linear combination of the components of $\mathbf{x}$
\[
m_{\theta}(t,s,\mathbf{x})=\Lambda(s)+\sum_{i=1}^{p}\mathbf{e}_1'e^{A(s-t)}\texttt{x}_i+
\int_t^s\mathbf{e}_1'e^{A(s-u)}\mathbf{e}_p\theta(u)\,du,
\]
so that 
\[
\frac{\partial}{\partial\texttt{x}_i}m_{\theta}(t,s,\mathbf{x})=\mathbf{e}_1'e^{A(s-t)}\mathbf{e}_i.
\]
\end{proof}
CDD futures prices depend nonlinearly on the vector $\mathbf{X}(t)$ 
which is included in the function $\Psi$. This fact makes difficult to 
derive analytic formulas for plain vanilla options (call options) which
are traded at the CME. To this aim, we recall some useful linearized formulas that allow to
approximate the CDD futures prices.
Let $t\leq \tau_1$, setting $\Psi(x)\approx x$ in~\eqref{FCDD-formula-new} reduces to
\begin{equation}
\label{FCDD-approx-x}
F_{\text{CDD}}(t,\tau_1,\tau_2,\texttt{x}_1,\ldots,\texttt{x}_p)_{\big\vert_{\mathbf{x}=\mathbf{X}(t)}}\approx
\Theta_{x}(t,\tau_1,\tau_2)+\mathbf{a}_{x}(t,\tau_1,\tau_2)\mathbf{X}(t)\,,
\end{equation}
where 
\begin{align*}
 &\mathbf{a}_{x}(t,\tau_1,\tau_2)=\int_{\tau_1}^{\tau_2}\mathbf{e}_1'\exp(A(s-t))\,ds\,,\\ 
 &\Theta_{x}(t,\tau_1,\tau_2)=\int_{\tau_1}^{\tau_2}
 c-\Lambda(s)\,ds+\int_{\tau_1}^{\tau_2}\int_t^s\mathbf{e}_1'\exp(A(s-u))\mathbf{e}_p\theta(u)\,du\,ds\,.
 \end{align*}
We can consider the first order Taylor approximation $\Psi(x)\approx\frac{1}{\sqrt{2\pi}}+\frac{1}{2}x$ instead, then \eqref{FCDD-formula-new} reduces to
\begin{equation}
\label{FCDD-approx-taylor}
F_{\text{CDD}}(t,\tau_1,\tau_2,\texttt{x}_1,\ldots,\texttt{x}_p)_{\big\vert_{\mathbf{x}=\mathbf{X}(t)}}\approx
\Theta_{\text{Taylor}}(t,\tau_1,\tau_2)+\mathbf{a}_{\text{Taylor}}(t,\tau_1,\tau_2)\mathbf{X}(t)\,,
\end{equation}
 where
\begin{align*}
\mathbf{a}_{\text{Taylor}}(t,\tau_1,\tau_2)&=\frac12\int_{\tau_1}^{\tau_2}\mathbf{e}_1'\exp(A(s-t))\,ds\,,\\ 
\Theta_{\text{Taylor}}(t,\tau_1,\tau_2)&=\int_{\tau_1}^{\tau_2}
\frac12(\Lambda(s)-c)+\frac{1}{\sqrt{2\pi}}\Sigma(t,s)\,ds \\
&+\frac12\int_{\tau_1}^{\tau_2}\int_t^s\mathbf{e}_1'\exp(A(s-u))\mathbf{e}_p\theta(u)\,du\,ds\,.
\end{align*}
We introduce a new notation that encompasses both approximated formulas for the CDD futures price. 
To this end let $t\leq\tau_1$, then 
\begin{equation}
\label{approx-FCDD-formula}
\widetilde{F}_{\text{CDD}}(t,\tau_1,\tau_2,\texttt{x}_1,\ldots,\texttt{x}_p)_{\big\vert_{\mathbf{x}=\mathbf{X}(t)}}
=\Theta(t,\tau_1,\tau_2)+\mathbf{a}(t,\tau_1,\tau_2)\mathbf{X}(t)\,,
\end{equation}
where $\Theta$ and $\mathbf{a}$ are generic notations for  
$\Theta_x$ and $\mathbf{a}_x$ or $\Theta_{\text{Taylor}}$ and $\mathbf{a}_{\text{Taylor}}$.
The next proposition provides the partial derivatives of the approximate CDD futures prices with respect to the components of $\mathbf{x}.$
\begin{proposition}
\label{prop:dappFCDD-ttau1tau2}
Let $t\leq \tau_1$, then it holds that 
\begin{equation}
\label{dappFCDD-ttau1tau2}
\frac{\partial}{\partial\texttt{x}_i}\widetilde{F}_{\text{CDD}}(t,\tau_1,\tau_2,\texttt{x}_1,\ldots,\texttt{x}_p)=
\mathbf{a}(t,\tau_1,\tau_2)\mathbf{e}_i\,,
\end{equation}
for $i=1,\ldots,p$ where $\mathbf{a}$ is the generic notation for $\mathbf{a}_{\text x}$ and $\mathbf{a}_{\text Taylor}$\,.
\end{proposition}
\begin{proof}
Rewrite \eqref{approx-FCDD-formula} in terms of the components of $\mathbf{x}$ as
$$
\widetilde{F}_{\text{CDD}}(t,\tau_1,\tau_2,\texttt{x}_1,\ldots,\texttt{x}_p)
=\Theta(t,\tau_1,\tau_2)+\sum_{i=1}^{p}\mathbf{a}(t,\tau_1,\tau_2)\mathbf{e}_i\texttt{x}_i\,,
$$
and differentiate with respect to the components $\texttt{x}_i$ for $i=1,\ldots,p$. 
\end{proof}
\noindent Observe that \eqref{dappFCDD-ttau1tau2} does not depend on $\mathbf{X}(t)$. 

We consider now CDD and HDD futures prices as the underlying to price call options. The arbitrage-free price for 
a call option with strike $K$ at exercise time $\tau$,
written on a CDD futures with measurement period $\left[\tau_1,\tau_2\right]$, 
with $\tau\leq\tau_1$ and for times $t\leq\tau_1$ is defined as 
\begin{equation}
\label{call}
C(t,\tau,\tau_1,\tau_2,K):=\e^{-r(\tau-t)}\mathbb{E}_Q\left[\max\left(F_{\text{CDD}}(\tau,\tau_1,\tau_2)-K,0\right)\,|\,\mathcal{F}_t\right]\,,
\end{equation}
where $r>0$ is the risk-free interest rate. All the stochasticity in the call option price is in
the term $\mathbf{e}_1'\e^{A(s-\tau)}\mathbf{X}(\tau)$ which is contained in the CDD futures price at exercise time
and more specifically in $m_{\theta}$. For $\tau\geq t$,
\begin{equation}
\label{X(tau)}
\mathbf{X}(\tau)=\e^{A(\tau-t)}\mathbf{X}(t)+\int_t^{\tau}\theta(s)\e^{A(\tau-s)}\mathbf{e}_p\,ds+\int_t^{\tau}\sigma(s)\e^{A(\tau-s)}\mathbf{e}_p\,dW(s)
\end{equation}
is a solution of the stochastic differential equation in \eqref{dynamics}. Therefore $\mathbf{e}_1'\e^{A(s-\tau)}\mathbf{X}(\tau)$ reduces to 
\begin{align}
\label{Z-rv}
\mathbf{e}_1'&\e^{A(s-\tau)}\mathbf{X}(\tau)\\
&=\mathbf{e}_1'\e^{A(s-t)}\mathbf{X}(t)+\int_{t}^{\tau}\theta(u)\mathbf{e}_1'\e^{A(s-u)}\mathbf{e}_p\,du+\int_t^{\tau}\sigma(u)\mathbf{e}_1'\e^{A(s-u)}\mathbf{e}_p\,dW(u),\nonumber
\end{align}
where the dependence on $\mathbf{X}(t)$ is explicited in $\mathbf{e}_1'\e^{A(s-t)}\mathbf{X}(t)$. We can use the same argument
as for the CDD futures price to rewritte the call option price as follows 
\begin{equation}
\label{call-new}
C(t,\tau,\tau_1,\tau_2,K,\texttt{x}_1,\ldots,\texttt{x}_p)_{\big\vert_{\mathbf{x}=\mathbf{X}(t)}}
=\e^{-r(\tau-t)}\mathbb{E}_Q\left[\max\left(F_{\text{CDD}}(\tau,\tau_1,\tau_2)-K,0\right)\,|\,\mathcal{F}_t\right].
\end{equation}

To study the sensitivity of the call option price with respcect to infinitesimal changes in the components of $\mathbf{X}(t)$ we have to consider also partial derivatives which is not an easy task, if possible.
To avoid differentiating the payoff $\max(F_{\text{CDD}}(\tau,\tau_1,\tau_2,\texttt{x}_1,\ldots,\texttt{x}_p)-K,0)$, 
it is known the density approach which moves the dependency of $\mathbf{X}(t)$ from the payoff to the required density function to compute the
conditional expectation, see Broadie and Glasserman \cite{B-G}. Next, we see that this method fails here because the payoff function is path-dependent
on $\mathbf{X}(t)$ from $\tau_1$ to $\tau_2$. Indeed, the payoff function contains the term $\mathbf{e}_1'\e^{A(s-\tau)}\mathbf{X}(\tau)$ in $m_{\theta}$ which 
depends on $\mathbf{X}(t)$ as we have seen in \eqref{Z-rv}. This fact makes not possible to perform the study of the sensitivity in call options written on CDD futures prices over a measurement period with respect to infinitesimal changes in $\mathbf{X}(t)$. 
The linearized CDD futures price as defined in \eqref{approx-FCDD-formula} makes possible to get 
an approximate call option price formula which is analytically treatable in the sense that 
approximation methods like Monte Carlo are not required.
This problem is thoroughly tackled by setting in \eqref{call-new}
the linearized CDD futures prices defined in \eqref{approx-FCDD-formula}, see Benth and Solanilla Blanco \cite{BenSol-A3} for a detailed explanation.
The approximate formula for the call option price then reduces to 
\begin{align}
\label{approx-call-tau1tau2}
\widetilde{C}(t,\tau,\tau_1,\tau_2,&K,\widetilde{F}_{\text{CDD}}(t,\tau_1,\tau_2,\texttt{x}_1,\ldots,\texttt{x}_p))_{\big\vert_{\mathbf{x}=\mathbf{X}(t)}}=\\
&\e^{-r(\tau-t)}S(t,\tau,\tau_1,\tau_2)\Psi\Big(\frac{d(t,\tau,\tau_1,\tau_2,\widetilde{F}_{\text{CDD}}(t,\tau_1,\tau_2))-K}{S(t,\tau,\tau_1,\tau_2)}\Big)\nonumber
\end{align}
with
$$
d(t,\tau,\tau_1,\tau_2,K,x)=x+\Theta(\tau,\tau_1,\tau_2)-\Theta(t,\tau_1,\tau_2)+
\int_t^{\tau}\theta(s)\mathbf{a}(s,\tau_1,\tau_2)\mathbf{e}_p\,ds
$$
and
$$
S^2(t,\tau,\tau_1,\tau_2)=\int_t^{\tau}\sigma^2(s)(\mathbf{a}(s,\tau_1,\tau_2)\mathbf{e}_p)^2\,ds\,.
$$
Observe that the approximate call option becomes explicitly dependent on the approximate futures price.
The next proposition provides the partial derivatives of the approximate call option price with respect to the components of $\mathbf{x}$.
\begin{proposition}
\label{pro:dappC-ttau1tau2}
Let $t\leq \tau_1$, then it holds that 
\begin{align*}
\big(\frac{\partial}{\partial x_i}\widetilde{C}(t,\tau,s,&K,\widetilde{F}_{\text{CDD}}(t,\tau_1,\tau_2,,\texttt{x}_1,\ldots,\texttt{x}_p))\big)_{\big\vert_{\mathbf{x}=\mathbf{X}(t)}}\\
&=e^{-r(\tau-t)}\Phi(\frac{d(t,\tau,s,\widetilde{F}_{\text CDD}(t,\tau_1,\tau_2))-K}{S(t,\tau,\tau_1,\tau_2)})\mathbf{a}(t,\tau_1,\tau_2)\mathbf{e}_{i}\,,
\end{align*} 
for $i=1,\ldots,p.$
\end{proposition}
\begin{proof}
The proof follows by taking partial derivatives in \eqref{approx-call-tau1tau2}. Take into account that  $\Psi'(x)=\Phi(x)$
and that the only component in the function $d$ dependent on the components of $\mathbf{X}(t)$ is the approximate
futures price whose partial derivatives are provided in Proposition.~\ref{prop:dappFCDD-ttau1tau2}.
\end{proof}
In the next section we simplify our setting to perform the study of sensitivity and consider futures prices with a measurement day and call options written on these.

\section{Sensitivity of CDD and HDD derivatives prices with a measurement day}\label{sec:teo-md}

In this Section we perform a complete study of the sensitivity of CDD and HDD futures prices with a measurement day and call options written on these to infinitesimal changes on the components of $\mathbf{X}(t)$.\\

The Fubini-Tonelli theorem, see e.g. Folland \cite{F} connects futures prices 
setting over a time period and futures prices with a delivery day running over a time period
as follows 
\begin{equation}
\label{Fubini}
F_{\text{CDD}}(t,\tau_1,\tau_2)
=\int_{\tau_1}^{\tau_2}\E_{Q}[\max(T(s)-c,0)|\mathcal{F}_t]\,ds=\int_{\tau_1}^{\tau_2}F_{\text{CDD}}(t,s)\,ds\,,
\end{equation}
for $t\leq s$. We see that $F_{\text{CDD}}(t,\tau_1,\tau_2)$ is expressed as the CDD futures price at time $t$ with a measurement day $s$, denoted $F_{\text{CDD}}(t,s)$, running over the time period $[\tau_1,\tau_2].$ Consequently, we deduce from \eqref{FCDD-formula} and \eqref{FHDD-formula} respectively that 
\[
F_{\text{CDD}}(t,s)=
\Sigma(t,s)\Psi\left(\frac{m_{\theta}(t,s,\mathbf{X}(t))-c}{\Sigma(t,s)}\right)\,
\]
and
\[
F_{\text{HDD}}(t,s)=
\Sigma(t,s)\Psi\left(\frac{c-m_{\theta}(t,s,\mathbf{X}(t))}{\Sigma(t,s)}\right)\,,
\]
where, for $\mathbf{x}\in\R^{p} $
\begin{align*}
m_{\theta}(t,s,\mathbf{x})&=\Lambda(s)+\mathbf{e}_1'e^{A(s-t)}\mathbf{x}+
\int_t^s\mathbf{e}_1'e^{A(s-u)}\mathbf{e}_p\theta(u)\,du \\
\Sigma^2(t,s) &=\int_t^s(\mathbf{e}_1'e^{A(s-u)}\mathbf{e}_p)^2\sigma^2(u)\,du\,.
\end{align*}
Recall that $\Psi(x)=x\Phi(x)+\Phi'(x)$, with $\Phi$ being the cumulative standard normal
distribution function. The same notation introduced in \eqref{FCDD-formula-new} can be used in this context, then
\[
F_{\text{CDD}}(t,s,\texttt{x}_1,\ldots,\texttt{x}_p)_{\big\vert_{\mathbf{x}=\mathbf{X}(t)}}=
\Sigma(t,s)\Psi\left(\frac{m_{\theta}(t,s,\mathbf{X}(t))-c}{\Sigma(t,s)}\right).
\]
The term $\mathbf{e}_1'\exp(A(s-t))\mathbf{X}(t)$ included in $m_{\theta}$
and rewritten as in \eqref{futures-term-tau1tau2} contains all the stochasticity 
of the futures prices and provides information about its evolution.
Note that when the time to measurement $x=s-t\rightarrow\infty$, 
the function $f_{i}(s-t)$ tends to zero for $i=1,\cdots,p$ since the real parts of 
the eigenvalues of the matrix $A$ are all strictly negative for having a stationary model.
Hence, at the long end we can say that 
the behavior of the future prices is not affected by this term.
But, if $x$ is approaching to zero, the term $\mathbf{e}_1'\exp(A(s-t))\mathbf{X}(t)$ 
is influenced for all the components of the $\mathbf{X}(t)$. Finally, for $x=0$, only the first component of $\mathbf{X}(t)$ 
takes part on the evolution of the futures prices. These arguments determine 
the evolution of futures prices at time $t>0$ when time to delivery is a day $s$,\,$s\geq t$.
In the case of futures prices with measurement period as presented in Section \ref{sec:teo-mp}, 
we have to take into account that $s$ is running over a measurement period.
In the next proposition we consider the partial derivatives of the CDD futures price with respect to the components of $\mathbf{x}$.
\begin{proposition}
\label{pro:dFCDD-ts}
Let $t\leq s$, it holds that 
\[
\big(\frac{\partial}{\partial \texttt{x}_i}F_{\text{CDD}}(t,s,\texttt{x}_1,\ldots,\texttt{x}_p)\big)_{\big\vert_{\mathbf{x}=\mathbf{X}(t)}}=
\Phi\left(\frac{m_{\theta}(t,s,\mathbf{X}(t))-c}{\Sigma(t,s)}\right)\mathbf{e}_1'\exp(A(s-t))\mathbf{e}_i
\]
for $i=1,\ldots,p$. 
\end{proposition}
\begin{proof}
The proof follows by applying the chain-rule. Consider that $\Psi'(x)=\Phi(x)$
and also take into account that $m_\theta$ can be written as a linear combination of the components of $\mathbf{x}$ as follows
\[
m_{\theta}(t,s,\mathbf{x})=\Lambda(s)+\sum_{i=1}^{p}\mathbf{e}_1'e^{A(s-t)}\texttt{x}_i+
\int_t^s\mathbf{e}_1'e^{A(s-u)}\mathbf{e}_p\theta(u)\,du,
\]
so that 
\[
\frac{\partial}{\partial \texttt{x}_i}m_{\theta}(t,s,\mathbf{x})=\mathbf{e}_1'e^{A(s-t)}\mathbf{e}_i.
\]
\end{proof}
Next, we also adapt the linearized formulas for CDD futures
prices presented before to our setting. To this end let $t\leq s$, 
formulas \eqref{FCDD-approx-x} and \eqref{FCDD-approx-taylor} reduce respectively to 
\begin{equation}
\label{FCDD-approx-x-ts}
F_{\text{CDD}}(t,s,\texttt{x}_1,\ldots,\texttt{x}_p)\approx
\Theta_{x}(t,s)+\mathbf{a}_{x}(t,s)\mathbf{x} \,,
\end{equation}
where 
\begin{align*}
&\mathbf{a}_{x}(t,s)=\mathbf{e}_1'\exp(A(s-t))\,\\ 
&\Theta_{x}(t,s)=
\Lambda(s)-c+\int_t^s\mathbf{e}_1'\exp(A(s-u))\mathbf{e}_p\theta(u)\,du,
\end{align*}
and 
\begin{equation}
\label{FCDD-approx-taylor-ts}
F_{\text{CDD}}(t,s,\texttt{x}_1,\ldots,\texttt{x}_p)\approx
\Theta_{\text{Taylor}}(t,s)+\mathbf{a}_{\text{Taylor}}(t,s)\mathbf{x}\,,
\end{equation} 
where
\begin{align*}
&\mathbf{a}_{\text{Taylor}}(t,s)=\frac12\mathbf{e}_1'\exp(A(s-t))\,,\\
&\Theta_{\text{Taylor}}(t,s)=\frac12(\Lambda(s)-c)+\frac{1}{\sqrt{2\pi}}\Sigma(t,s)+\frac12\int_t^s\mathbf{e}_1'\exp(A(s-u))\mathbf{e}_p\theta(u)\,du\,.
\end{align*}
We provide also the new notation that encompasses 
both approximate CDD futures prices formulas. To this end let $t\leq s$, then
\begin{equation}
\label{approx-FCDD-formula-ts}
\big(\widetilde{F}_{\text{CDD}}(t,s,\texttt{x}_1,\ldots ,\texttt{x}_p)\big)_{\big\vert_{\mathbf{x}=\mathbf{X}(t)}}=\Theta(t,s)+\mathbf{a}(t,s)\mathbf{X}(t)
\end{equation}
where $\Theta$ and $\mathbf{a}$ are generic notations for  
$\Theta_x$ and $\mathbf{a}_x$ or $\Theta_{\text{Taylor}}$ and $\mathbf{a}_{\text{Taylor}}$.
The next proposition provides the partial derivatives of the approximate CDD futures prices with respect to the components of $\mathbf{x}$.
\begin{proposition}
\label{pro:dappFCDD-ts}
Let $t\leq s$, then it holds that 
\begin{equation}
 \label{dappFCDD-ts}
\frac{\partial}{\partial \texttt{x}_i}\widetilde{F}_{\text{CDD}}(t,s,\texttt{x}_1,\ldots,\texttt{x}_p)=
\mathbf{a}(t,s)\mathbf{e}_i
\end{equation}
for $i=1,\ldots,p$, where $\mathbf{a}$ is the generic notation for $\mathbf{a}_{\text x}$ and $\mathbf{a}_{\text Taylor}$.
\end{proposition}
\begin{proof}
Rewrite \eqref{approx-FCDD-formula-ts} in terms of the components of $\mathbf{x}$ as
$$
\widetilde{F}_{\text{CDD}}(t,s,\texttt{x}_1,\ldots,\texttt{x}_p)
=\Theta(t,s)+\sum_{i=1}^{p}\mathbf{a}(t,s)\mathbf{e}_i\texttt{x}_i\,,
$$
and differentiate with respect to the components $\texttt{x}_i$ for $i=1,\ldots,p$.
\end{proof}
\noindent Observe that unlike the result in Proposition.~$\ref{pro:dFCDD-ts}$, here we loose the dependency on $\mathbf{X}(t)$.

The arbitrage-free price for a call option with strike $K$ at exercise time $\tau$,
written on a CDD futures with measurement day $s$, for a time $t$
with $t\leq\tau\leq s$ is defined as 
\begin{equation}
\label{call-ts}
C(t,\tau,s,K):=\e^{-r(\tau-t)}\mathbb{E}_Q\left[\max\left(F_{\text{CDD}}(\tau,s)-K,0\right)\,|\,\mathcal{F}_t\right]\,,
\end{equation}
where $r>0$ is the risk-free interest rate. For our purposes and making use of the same argument as in Section \ref{sec:teo-mp} we can rewritte \eqref{call-ts} as
\begin{equation}
\label{call-new-ts}
C(t,\tau,s,K,\texttt{x}_1,\ldots,\texttt{x}_p)_{\big\vert_{\mathbf{x}=\mathbf{X}(t)}}
:=\e^{-r(\tau-t)}\mathbb{E}_Q\left[\max\left(F_{\text{CDD}}(\tau,s)-K,0\right)\,|\,\mathcal{F}_t\right].
\end{equation}
Next, we see that the density approach here works well as the payoff function of the call option price is 
not path-dependent on $\mathbf{X}(t)$ over a time period. In the next Proposition we present then the partial derivatives of the call option price
with respect to the components of $\mathbf{x}$. 
\begin{proposition}
\label{pro:dC-ts}
Let $t\leq s$, then it holds that 
\[
\big(\frac{\partial}{\partial\texttt{x}_i}C(t,\tau,s,K,\texttt{x}_1,\ldots,\texttt{x}_p)\big)_{\big\vert_{\mathbf{x}=\mathbf{X}(t)}}=
e^{-r(\tau-t)}\E_{Q}\left[g(Z,t,s,\tau,\mathbf{X}(t))\right]
\]
for $i=1,\ldots,p$, where for $\mathbf{x}\in\R^p$
$$g(Z,t,s,\tau,\mathbf{x})=\max(F_{\text CDD_{,Z}}(\tau,s)-K,0)\left(\frac{Z-\widetilde{m}_\theta(t,s,\tau,\mathbf{x})}{\widetilde{\Sigma}^2(t,s,\tau)}\right)\mathbf{e}_1'\exp(A(s-t))\mathbf{e}_i\,.$$
$Z=\mathbf{e}_1'\exp(A(s-\tau))\mathbf{X}(\tau)$ is a normally distributed random variable and  
\begin{align*}
\widetilde{m}_\theta(t,s,\tau,\mathbf{x})&=\mathbf{e}_1'\exp(A(s-t))\mathbf{x}
+\int_{t}^{\tau}\mathbf{e}_1'\exp(A(s-u))\mathbf{e}_p\theta(u)\,du\,\\
\widetilde{\Sigma}^2(t,s,\tau)&=\int_{t}^{\tau}(\mathbf{e}_1'\exp(A(s-u))\mathbf{e}_p)^2\sigma^{2}(u)\,du
\end{align*}
are the mean and the variance of $Z$ conditioned on $\mathbf{X}(t)$, respectively. 
\end{proposition}
\begin{proof}
The random variable $\mathbf{e}_1'\exp(A(s-\tau))\mathbf{X}(\tau)=Z$ included in $m_{\theta}$
is normally distributed and 
\[
\tilde{m}_{\theta}(t,s,\tau,\mathbf{x})=\mathbf{e}_1'\e^{A(s-t)}\mathbf{x}+\int_{t}^{\tau}\theta(u)\mathbf{e}_1'\e^{A(s-u)}\mathbf{e}_p\,du
\]
and
\[
\tilde{\Sigma}^{2}(t,s,\tau) =\int_{t}^{\tau}\sigma^2(u)(\mathbf{e}_1'\exp(A(s-u))\mathbf{e}_p)^2\,du\,
\]
are the mean and the variance of $Z$, respectively, conditioned on 
$\mathbf{X}(t).$ The probability density function of $Z$ is then
\[
p_{Z}(z,t,s,\tau,\mathbf{x})=\frac{1}{\sqrt{2\pi}\tilde{\Sigma}(t,s,\tau)}\exp\Big(-\frac{1}{2}\Big(\frac{z-\tilde{m}(t,s,\tau,\mathbf{x})}
{\tilde\Sigma(t,s,\tau)}\Big)^{2}\Big)\,,
\]
 We see that we have moved the dependency of $\mathbf{X}(t)$ contained in 
$Z$ from the payoff function to the required density function to compute the
the conditional expectation as follows
\begin{align*}
\big(\frac{\partial}{\partial\texttt{x}_i}C&(t,\tau,s,K,\texttt{x}_1,\ldots,\texttt{x}_p)\big)_{\big\vert_{\mathbf{x}=\mathbf{X}(t)}}\\
&=e^{-r(\tau-t)}\int_{\R}\max(F_{\text CDD_{,z}}(\tau,s)-K,0)\big(\frac{\partial}{\partial\texttt{x}_i}P_{Z}(z,t,s,\tau,\mathbf{x})\big)\,dz_{\big\vert_{\mathbf{x}=\mathbf{X}(t)}}\,\\
&=e^{-r(\tau-t)}\int_{\R}\max(F_{\text CDD_{,z}}(\tau,s)-K,0)\big(\frac{z-\widetilde{m}_{\theta}(t,s,\tau,\mathbf{x})}{\widetilde{\Sigma}^2(t,s,\tau)}\big)\mathbf{e}_1'\e^{A(s-t)}\mathbf{e}_iP_{Z}(z,t,s,\tau,\mathbf{x})\,dz_{\big\vert_{\mathbf{x}=\mathbf{X}(t)}}\,\\
&=\E_{Q}\left[\max(F_{\text CDD_{,Z}}(\tau,s)-K,0)\big(\frac{Z-\widetilde{m}_{\theta}(t,s,\tau,\mathbf{X}(t))}{\widetilde{\Sigma}^2(t,s,\tau)}\big)\mathbf{e}_1'\e^{A(s-t)}\mathbf{e}_i\,|\,\mathcal{F}_t \right].
\end{align*}
\end{proof}
Next, we adapt to our setting the approximate call option price formula in \eqref{approx-call-tau1tau2}  
which reduces to 
\begin{align}
\label{approx-call-ts}
\widetilde{C}(t,\tau,s,&K,\widetilde{F}_{\text{CDD}}(t,s,\texttt{x}_1,\ldots,\texttt{x}_p))_{\big\vert_{\mathbf{x}=\mathbf{X}(t)}}\\
&=\e^{-r(\tau-t)}S(t,\tau,s)\Psi\Big(\frac{d(t,\tau,s,\widetilde{F}_{\text{CDD}}(t,s))-K}{S(t,\tau,s)}\Big)\nonumber
\end{align}
with
$$
d(t,\tau,s,K,x)=x+\Theta(\tau,s)-\Theta(t,s)+
\int_{t}^{\tau}\theta(u)\mathbf{a}(u,s)\mathbf{e}_p\,du
$$
and
$$
S^2(t,\tau,s)=\int_t^{\tau}\sigma^2(u)(\mathbf{a}(u,s)\mathbf{e}_p)^2\,du\,.
$$

We end up this Section with a result for the partial derivatives of the approximate call option price
with respect to the components of $\mathbf{x}$.
\begin{proposition}
\label{pro:dappC-ts}
Let $t\leq s$, then it holds that 
\begin{align*}
\big(\frac{\partial}{\partial\texttt{x}_i}\widetilde{C}(t,\tau,s,&K,\widetilde{F}_{\text{CDD}}(t,s,\texttt{x}_1,\ldots,\texttt{x}_p))\big)_{\big\vert_{\mathbf{x}=\mathbf{X}(t)}}=
e^{-r(\tau-t)}\Phi(\frac{d(t,\tau,s,\widetilde{F}_{\text CDD}(t,s))-K}{S(t,\tau,s)})\mathbf{a}(t,s)\mathbf{e}_i.
\end{align*}   
for $i=1,\ldots,p.$
\end{proposition}
\begin{proof}
The proof follows by taking partial derivatives in \eqref{approx-call-ts}. Take into account that  $\Psi'(x)=\Phi(x)$
and that the only component in function $d$ dependent on the components of $\mathbf{X}(t)$ is the approximate
futures price whose partial derivatives are provided in Proposition.~\ref{pro:dappFCDD-ts}.
\end{proof}

\section{Empirics}\label{sec:empirics}

Consider the stationary CAR($3$)-process obtained to model the 
temperature data from New York in Benth and Solanilla Blanco \cite{BenSol-A3}
which is defined with the following mean reverting matrix $A$,
$$
A=\begin{pmatrix}
0 & 1 & 0 \\
0 & 0 & 1 \\
-0.3364  & -1.6105 & -2.1618
\end{pmatrix}
$$
and a constant volatility, $\sigma=5.25$. The function $\Sigma^2$ which defines 
the CDD and HDD futures price now reduces to  
$$
\Sigma^2(t,s):=\Sigma^2(s-t)=\sigma^2\int_0^{s-t}(\mathbf{e}_1'\e^{Au}\mathbf{e}_p)^2\,du\,.
$$
Furthermore we choose to work with $\theta=0$ and fix the measurement day as August 1st, 2011.
We focus our empirical study on CDD futures prices with a measurement day being August 1st, 2011 
and call options written on these.
We include also a final section with some empirics on CDD futures prices with a delivery period being 
August 2011. We choose to work with delivery in August 2011, whether it is a particular day 
or all the month, as it was proved in Benth and Solanilla Blanco \cite{BenSol-A3} that
the approximative formula for the CDD futures price worked well in this time period.  

\begin{figure} [h!]
\includegraphics[height=2in, width=6in]{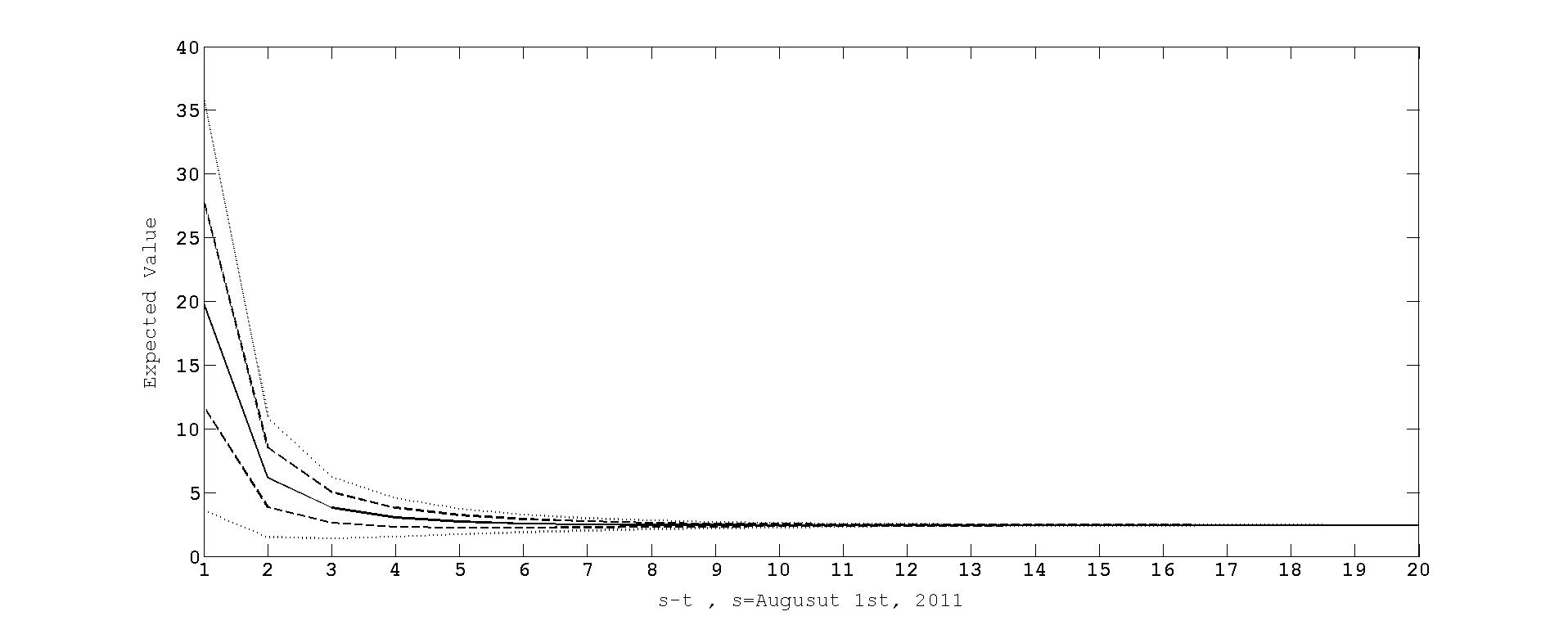}
\caption{Expected value of $Z(t,s)$ (complete line) as a function of $s-t$ with measurement day $s$ being August 1st, 2011. In addition, we have inserted the bounds for $\pm1$ std (slashed line) and $\pm2$ std (dotted line).}
\label{fig:Expectedvalue}
\end{figure}

To analyze the sensitivity of CDD futures prices with measurement as August 1st 2011 we consider
Proposition. \ref{pro:dFCDD-ts} and Proposition. \ref{pro:dappFCDD-ts} in Section \ref{sec:teo-md}.
In this context, the random variable $\Phi(m_{\theta}(t,s,\mathbf{X}(t))-c)/\Sigma(s-t))$ 
which makes the difference between the result provided in Proposition~\ref{pro:dFCDD-ts} 
and Proposition~\ref{pro:dappFCDD-ts} can be rewritten as $\Phi(Z(t,s))$ where  
\begin{equation}
\label{Z(t,s)}
Z(t,s):=(m_{\theta}(t,s,\mathbf{X}(t))-c)/\Sigma(s-t),
\end{equation}
and $s-t$ is the time to maturity. In Benth and Solanilla Blanco \cite{BenSol-A3}, 
a more general study with $\theta$ being a time dependent function concludes that
when $s-t\downarrow 0$ the expected value of $Z(t,s)$ tends to $\pm\infty$ and the 
variance $\infty$. Such a case indicates too much dispersion.
On the other hand, when $s-t\rightarrow\infty$, the variance of $Z(t,s)$ tends to zero, then 
$Z(t,s)\approx\E_{Q}(Z(t,s))$. Figure. \ref{fig:Expectedvalue} shows the tendency $Z(t,s)$
when the measurement day $s$ is August 1st,  2011.
\begin{figure} [h!]
\includegraphics[height=2in, width=6in]{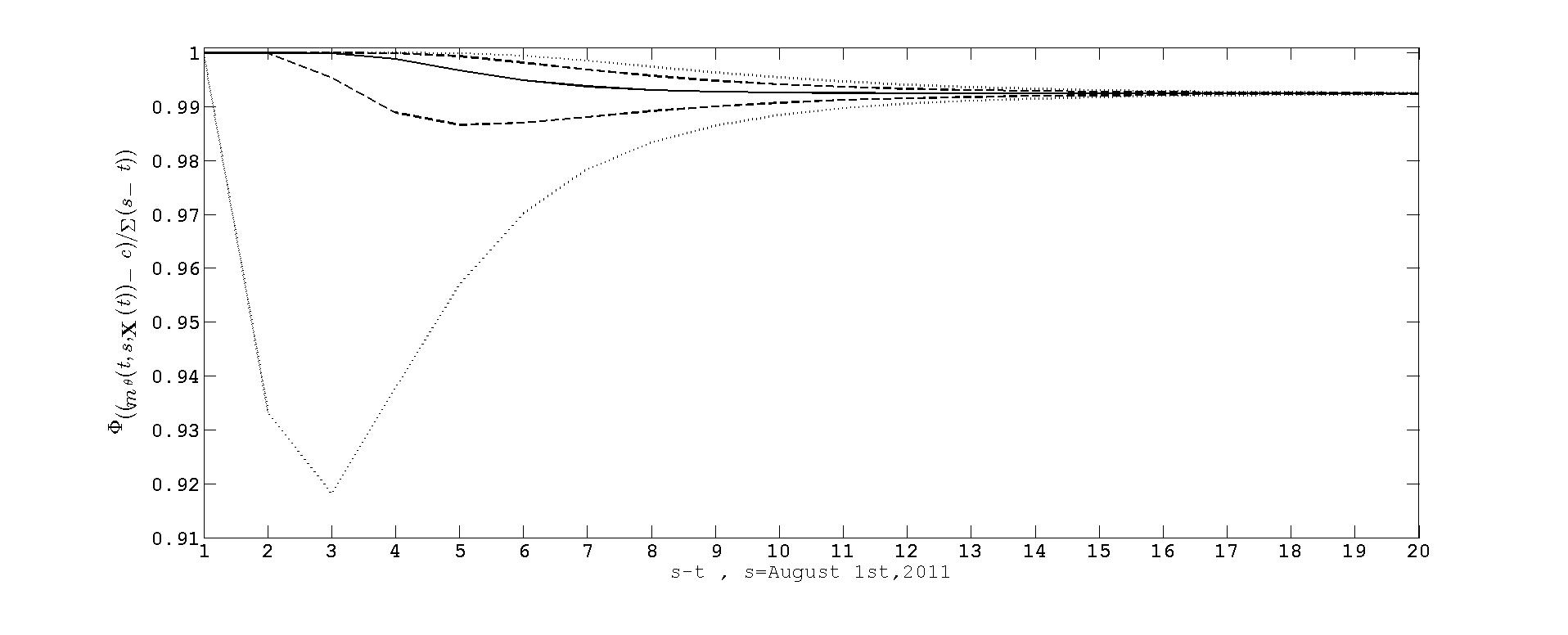}
\caption{$\Phi(\E_{Q}(Z(t,s)))$ (complete line) as a function of $s-t$, where we have chosen $s$ as August 1st, 2011. In addition, we have inserted $\Phi(\E_{Q}(Z(t,s))\pm i\,\text{std})$ for $i=1$ (slashed line) and $i=2$ (dotted line).}
\label{fig:Phi}
\end{figure}
Figure.~\ref{fig:Phi} is the result of applying the function $\Phi$ to the plot in Figure.~\ref{fig:Expectedvalue}. We see that 
$\Phi(Z(t,s))=1$ for $s-t=1,2$. From $s-t=3$ to $s-t=11$ there is a small decay and finally $\Phi(Z(t,s))$ stabilizes at 0.9924 for $s-t\geq 12$.
The values between the dashed lines are more probable than the ones in between the dotted lines.
We deduce that 
$$
\big(\frac{\partial}{\partial\texttt{x}_i}F_{\text{CDD}}(t,s,\texttt{x}_1,\texttt{x}_2,\texttt{x}_3)\big)_{\vert_{\mathbf{x}=\mathbf{X}(t)}}
\leq
\frac{\partial}{\partial\texttt{x}_i}\widetilde{F}_{\text CDD}(t,s,\texttt{x}_1,\texttt{x}_2,\texttt{x}_3),
$$
for $i=1,\ldots,3$\,, i.e., the approximate CDD futures prices are more sensitive to changes in the components of $\mathbf{X}(t)$ than the CDD futures prices. 

Recall from Benth and Solanilla Blanco \cite{BenSol-Japan} that $\mathbf{X}(t)$ 
contains the deseasonalized temperature $Y(t)$ and its derivatives
up to order $p-1$. In our setting then, $\mathbf{X}(t)$ reduces to
\begin{equation}
\label{X}
\mathbf{X}(t)=\left(\begin{array}{c} Y(t) \\
Y'(t)\\
Y''(t)\end{array}\right),
\end{equation}\
where $Y'(t)$ and $Y''(t)$ are respectively the slope and the curvature of $Y(t)$, respectively. 
\begin{figure} [h!]
\includegraphics[height=2in, width=6in]{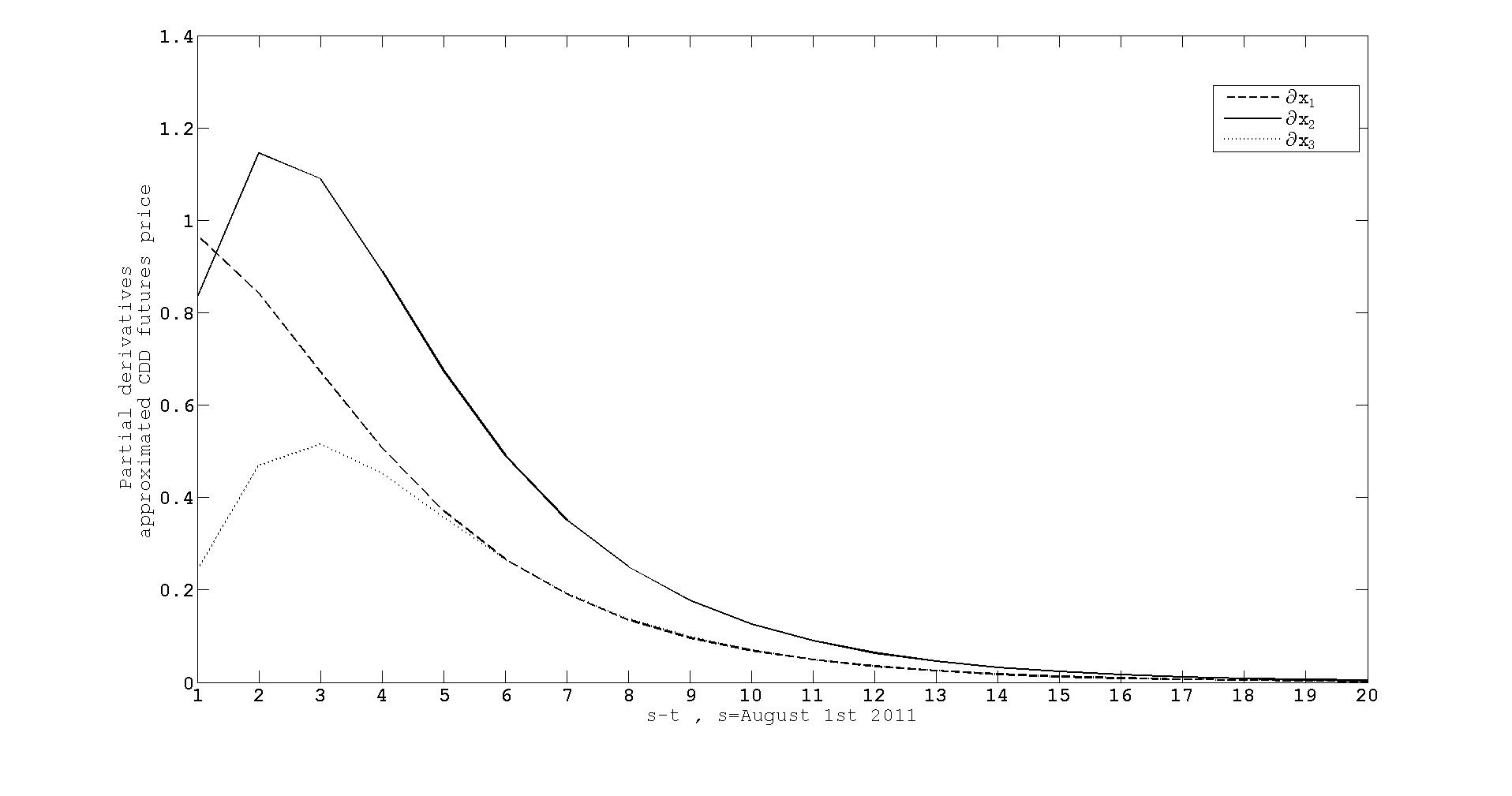}
\caption{$\partial\widetilde{F}_{\text{CDD}}(t,s,\texttt{x}_1,\ldots,\texttt{x}_p)/\partial\texttt{x}_i$
for\,$i=1,2,3$ as a function of $s-t$ with measurement day $s$ being August 1st, 2011.}
\label{fig:dappFCDD}
\end{figure}

Figure.~\ref{fig:dappFCDD} shows the partial derivatives of the approximate CDD futures price with respect the coordinates of $\mathbf{x}$ as presented in Proposition. \ref{pro:dappFCDD-ts}. 
The x-axis considers the time to maturity, $s-t$, where the measurement day $s$ is August 1st, 2011. The y-axis shows the different partial derivatives.
Firstly, we observe that the partial derivatives of the approximate CDD futures price are positive.
We see that at time to maturity $s-t=1$ any perturbation in the component $\texttt{x}_1$ affects the tendency 
of the approximate CDD futures price more than in the components $\texttt{x}_2$ and $\texttt{x}_3$.
However, when time to maturity increases the contribution of $\texttt{x}_1$ decreases gradually.
From $s-t=1$ to $s-t=2$ the contribution of $\texttt{x}_2$ increases
to the extent that at time to maturity $s-t=2$ perturbations in $x_2$ dominate the evolution of the approximate CDD futures price.  
From $s-t>2$ the contribution of $\texttt{x}_2$ decreases gradually but it remains always above $\texttt{x}_1$.
The contribution of $\texttt{x}_3$ increases from $s-t=1$ to $s-t=3$. For bigger times to maturity it decreases gradually. 
We point out that variations in $\texttt{x}_3$ always contribute less than variations in $\texttt{x}_1$ or $\texttt{x}_2$.
At the long end, small variations in any component hardly affect the tendency of the approximate CDD futures price. 
This fact makes sense as the term $\mathbf{e}_1e^{A(s-t)}\mathbf{x}$, which is dependent on
the coordinates of $\mathbf{x}$, tends to zero at the long end.
The partial derivatives of the CDD futures price with respect the coordinates of $\mathbf{x}$, as presented in Proposition. \ref{pro:dFCDD-ts}, depend on $\mathbf{X}(t)$.   
The first component $Y(t)=T(t)-\Lambda(t)$ corresponds to the deseasonalized temperature.
We approximate the derivatives of $Y(t)$ with backward finite differences.
Hence $Y'(t)\approx Y(t)-Y(t-1)$ is approximated by  
the difference between the deseasonalized temperature at times $t$ and $t-1$.  
$Y''(t)\approx Y(t)-2Y(t-1)+Y(t-2)$ is approximated by a linear combination of the 
deseasonalized temperatures at times $t$ and the two prior times $t-1$ and $t-2$. 
Finally, we get the following relation between the temperature and the seasonal
function:

\begin{align}
\label{T-Lambda}
\begin{pmatrix}\nonumber
T(t) \\
T(t-1)\\
T(t-2) 
\end{pmatrix} 
&\approx
\begin{pmatrix}
\Lambda(t)+Y(t)\\
\Lambda(t-1)+ Y(t)-Y'(t) \\
 \Lambda(t-2)+Y''(t)-2Y'(t-1)+Y(t)
\end{pmatrix}_{\Big\vert_{\mathbf{X(t)}=\mathbf{x}}}\\
&=
\begin{pmatrix}
\Lambda(t)+\texttt{x}_1\\
\Lambda(t-1)+ \texttt{x}_1-\texttt{x}_2 \\
 \Lambda(t-2)+\texttt{x}_3-2\texttt{x}_2+\texttt{x}_1
\end{pmatrix}.
\end{align}
Observe that given a fixed $\mathbf{X}(t)$ with $0\leq t\leq s$, the temperature at time $t$
is approximately $\texttt{x}_1$ degrees above the seasonal mean function and one and tho days prior to $t$, it is  
approximately $\texttt{x}_1-\texttt{x}_2$ and  
$\texttt{x}_3-2\texttt{x}_2+\texttt{x}_1$ degrees above the seasonal mean function, respectively.

Consider $\mathbf{X}(t)=\mathbf{0}$ where $\mathbf{0}$ is the null vector in $\R^3$. The partial derivatives of the CDD futures price are completely deterministic. 
By the relation between the temperature and the seasonal function established in \eqref{T-Lambda}, for this particular case the temperature for the time $t$ and the two prior times $t-1$ and $t-2$ is approximately 
the seasonal mean function. 

 \begin{figure} [h!]
\includegraphics[height=2in, width=6in]{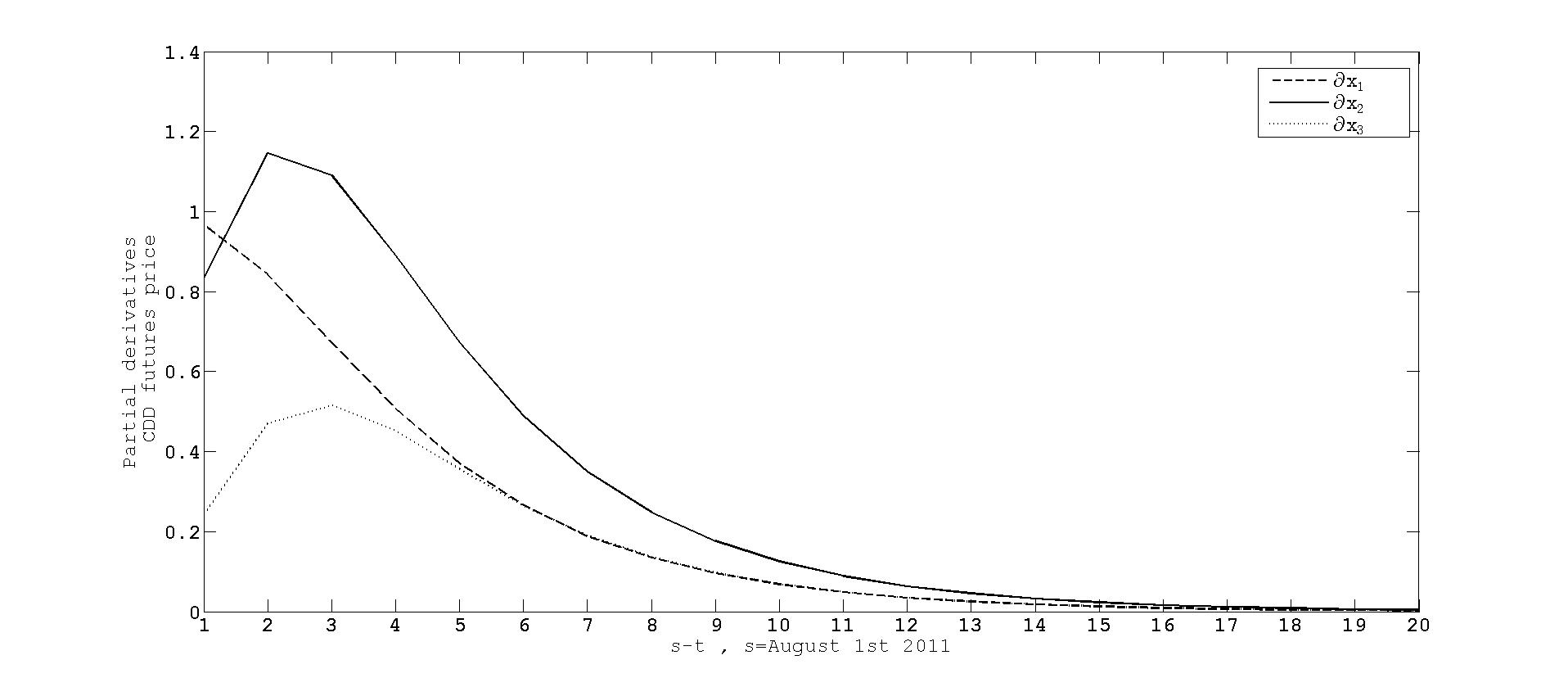}
\caption{$(\partial F_{\text{CDD}}(t,s,\texttt{x}_1,\ldots,\texttt{x}_p)/\partial\texttt{x}_i)_{\vert_{\mathbf{x}=\mathbf{0}}}$ 
as a function of $s-t$ with measurement day $s$ being August 1st, 2011.}
\label{fig:dFCDD000}
\end{figure}

Figure.~\ref{fig:dFCDD000} shows the partial derivatives of the CDD futures price with respect the coordinates of $\mathbf{x}$ derived in Proposition \ref{pro:dFCDD-ts}.
The x-axis considers the time to maturity, $s-t$, where the measurement day $s$ is August 1st, 2011. The y-axis shows the
different partial derivatives. 

Observe that at first sight Figure.~\ref{fig:dappFCDD} and Figure.~\ref{fig:dFCDD000} seem to coincide. 
Indeed, Figure. \ref{fig:relerror-000-A4} below shows that the relative error between them is less than 1$\%$ entirely.
\begin{figure} [h!]
\begin{center}
\includegraphics[height=2in, width=6in]{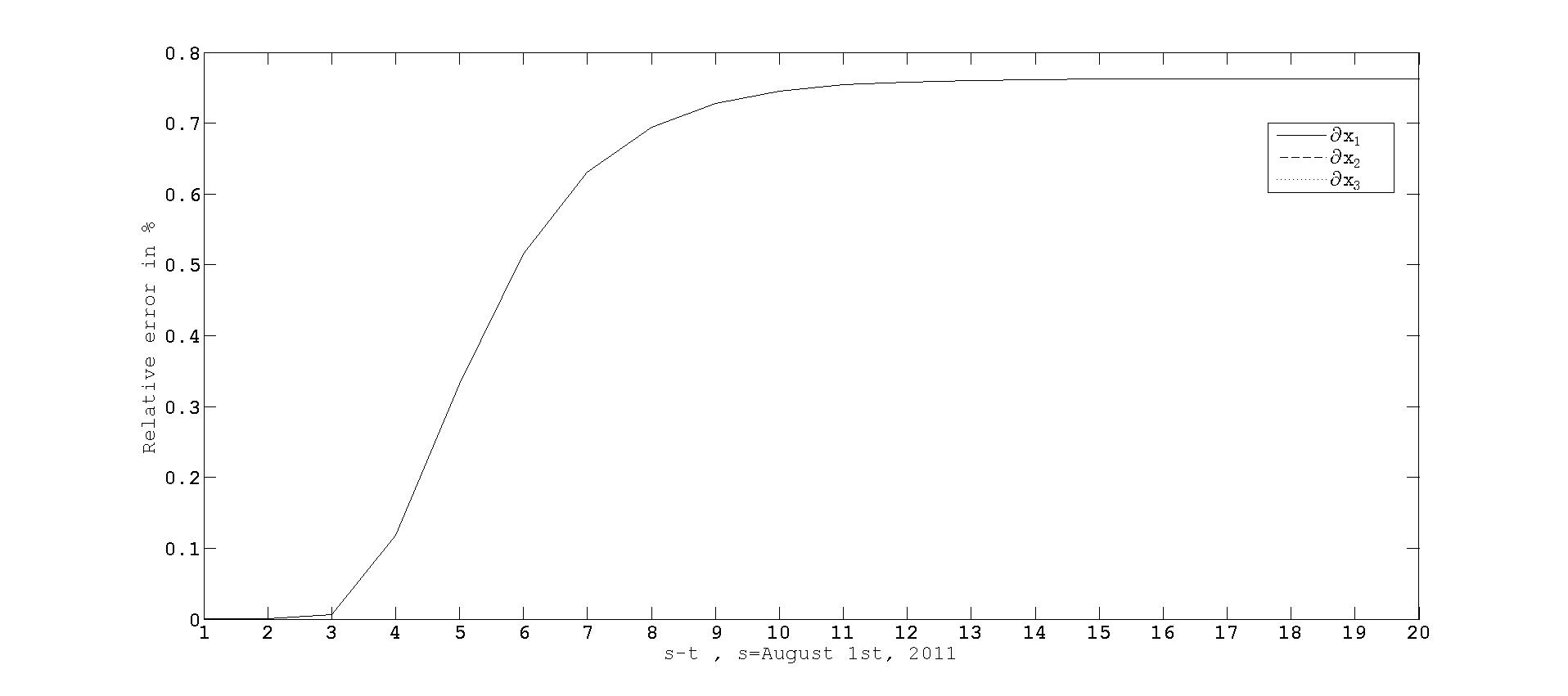}
\caption[Relative error in percent between the partial derivatives of the CDD futures price at $\mathbf{x}=\mathbf{e}_1\in\R^3$ and the 
partial derivatives of the approximate CDD futures price.]
{The relative error in percent between $\partial\widetilde{F}_{\text{CDD}}(t,s,\texttt{x}_1,\ldots,\texttt{x}_p)/\partial\texttt{x}_i$
and $(\partial F_{\text{CDD}}(t,s,\texttt{x}_1,\ldots,\texttt{x}_p)/\partial\texttt{x}_i)_{\vert_{\mathbf{x}=\mathbf{0}}}$ for\,$i=1,2,3$
as a function of $s-t$ with measurement day $s$ being August 1st, 2011.}
\label{fig:relerror-000-A4}
\end{center}
\end{figure}

Consider also the case where $\mathbf{X}(t)=\mathbf{e}_{k}$ is the $k$th canonical basis vector in $\R^3$ for $k=1,\ldots,3$. 
For $\mathbf{X}(t)=\mathbf{e}_{1}$  
the temperature at the present time $t$ and at the two consecutive prior times to $t$, say $t-1$ and $t-2$, is approximately one degree above the seasonal mean function. 
For $\mathbf{X}(t)=\mathbf{e}_{2}$, the temperature is close to the seasonal mean at present time $t$ and
it is approximately one and two degrees below the seasonal mean at times $t-1$ and $t-2$, respectively.
Finally for $\mathbf{X}(t)=\mathbf{e}_{3}$ the temperature is close to the seasonal mean at present time $t$ 
and one prior time, but two days prior to $t$ it is nearly one degree above its seasonal mean.  The partial derivatives
of the CDD futures price evaluated at the canonical basis vectors in $\R^3$ also behave in a similar way to the partial
derivatives on the approximate CDD futures price. Figure.~\ref{fig:relerror-100-A4} below shows that, for the case $\mathbf{X}(t)=\mathbf{e}_1$,  
the relative error between the partial derivatives of the approximate CDD futures price and the CDD futures price
is also less than 1\% entirely.
\begin{figure} [h!]
\begin{center}
\includegraphics[height=2in, width=6in]{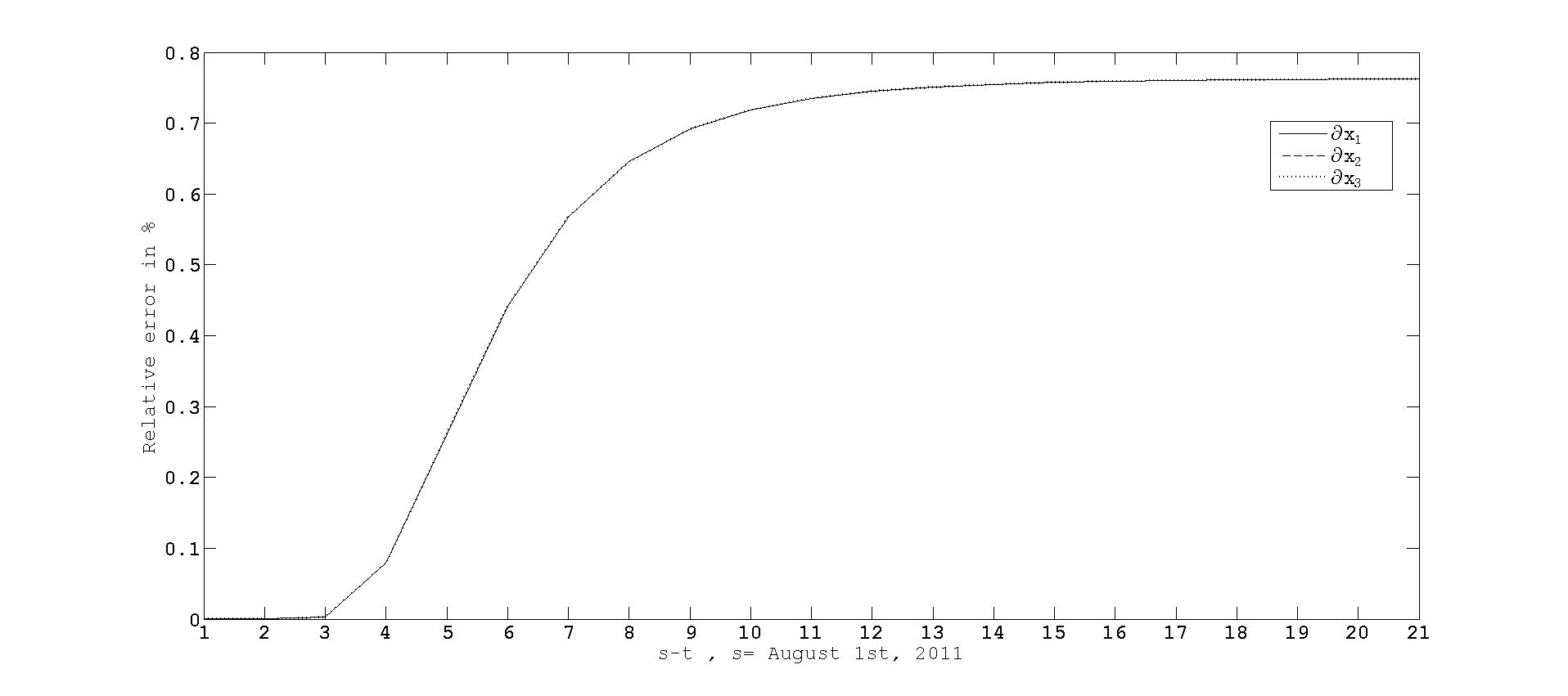}
\caption[$(\partial F_{\text{CDD}}(t,s,\texttt{x}_1,\ldots,\texttt{x}_p)/\partial\texttt{x}_i)_{\vert_{\mathbf{x}=\mathbf{0}}}$ 
for\,$i=1,2,3$  as a function of $s-t$ where $s$ is August 1st, 2011.]
{The relative error in percent between $\partial\widetilde{F}_{\text{CDD}}(t,s,\texttt{x}_1,\ldots,\texttt{x}_p)/\partial\texttt{x}_i$
and $(\partial F_{\text{CDD}}(t,s,\texttt{x}_1,\ldots,\texttt{x}_p)/\partial\texttt{x}_i)_{\vert_{\mathbf{x}=\mathbf{e}_1}}$ for\,$i=1,2,3$
as a function of $s-t$ with measurement day $s$ being August 1st, 2011.}
\label{fig:relerror-100-A4}
\end{center}
\end{figure}

\begin{figure}[h!]
\begin{center}
\includegraphics[height=2in, width=6in]{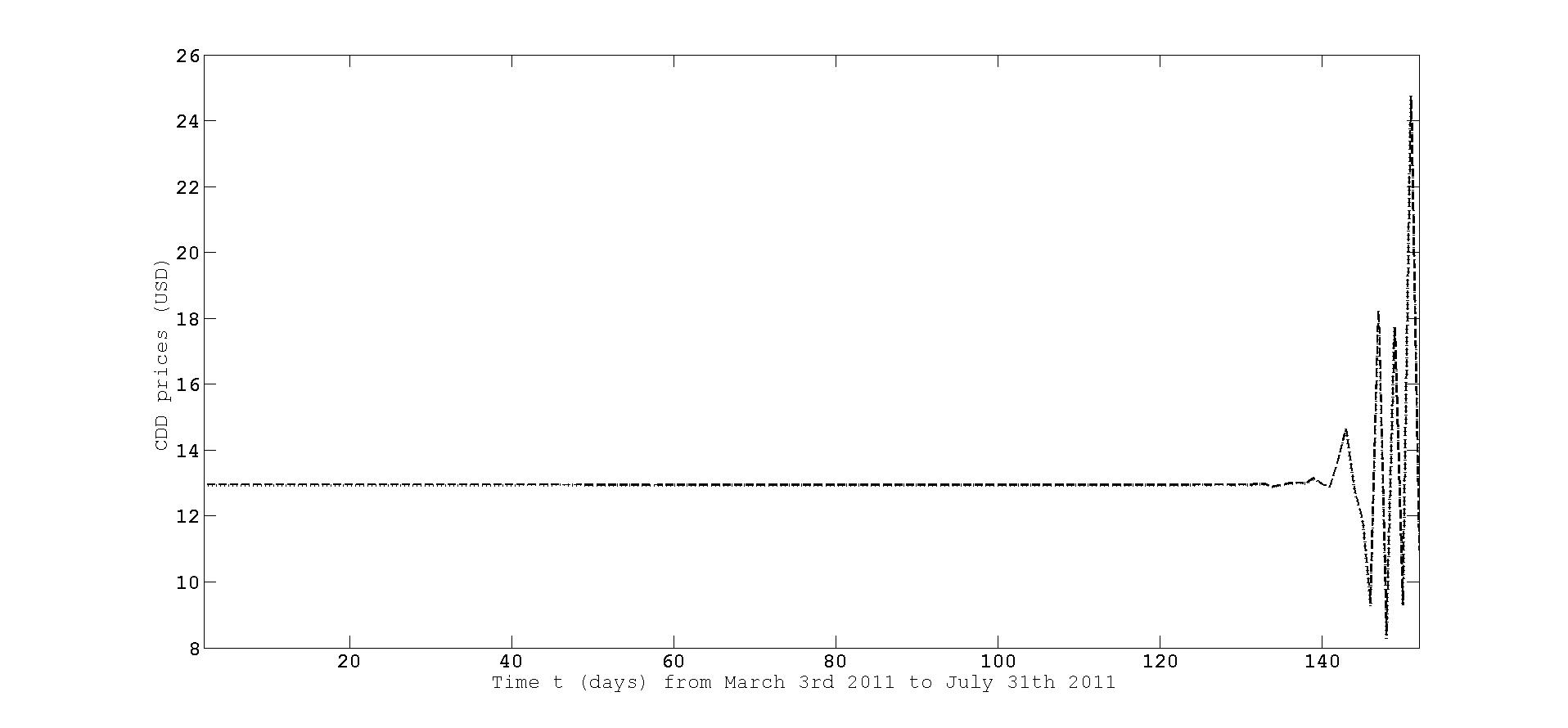}
\caption{Forward prices and approximate forward prices for CDD contracts from March 3rd, 2011 to July 31th, 2011 
with measurement day August 2nd, 2011.}
\label{fig:K-determination}
\end{center}
\end{figure}

We proceed now with the analysis of call option prices written on CDD futures prices and call option prices 
written on the approximate CDD futures prices. To do so, we consider the results in 
Propositons~\ref{pro:dC-ts} and \ref{pro:dappC-ts}. Observe that in both results there is dependency on
$\mathbf{X}(t)$. We focus on an at-the-money call option prices. In view of Figure.~\ref{fig:K-determination}
we fix the strike price $K$ being $K=13$.

For the study of the sensitivity of call option prices, we restrict our attention to the cases $\mathbf{X}(t)=\mathbf{0}$ and $\mathbf{X}(t)=\mathbf{e}_1$.
Next, we show the plots with the partial derivatives of the (approximate) call option prices
with respect to the coordinates of $\mathbf{x}$ derived in Proposition. \ref{pro:dC-ts} and Proposition. \ref{pro:dappC-ts}.

\begin{figure}[h!]
\begin{center}
\includegraphics[height=2in, width=6in]{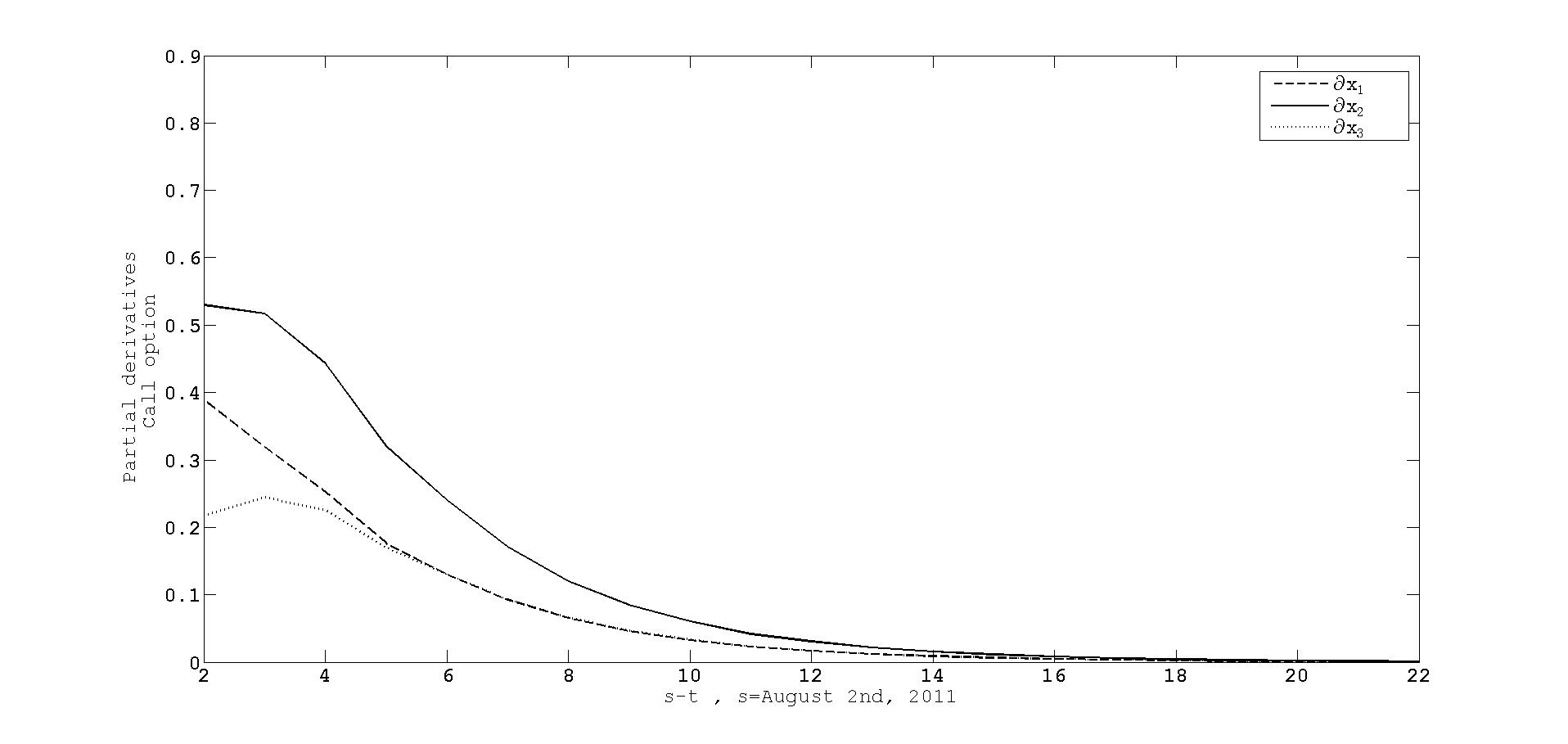}
\caption{$(\partial C(t,\tau,s,\texttt{x}_1,\ldots,\texttt{x}_p)/\partial\texttt{x}_i)_{\vert_{\mathbf{x}=\mathbf{0}}}$ for $i=1,2,3$
with exercise time $\tau$ being August 1st, 2011 and measurement day $s$ August 2nd, 2011.}
\label{fig:dC000}
\end{center}
\end{figure}

The x-axis considers the time to maturity, $s-t$, where $s$ is August 2nd, 2011. We have fixed the exercise time $\tau$ being August 1st, 2011.
The y-axis shows the different partial derivatives of the call option price when $\mathbf{X}(t)=\mathbf{0}$. 
Firstly, we observe that all the partial derivatives of the call option are positive. 
We see that for all times to maturity the call option price is more 
sensitive to any infinitesimal change in the component $\texttt{x}_2$, followed by $\texttt{x}_1$ and $\texttt{x}_3$. 
This tendency follows as time to maturity increases.
At the long end, small variations in any component hardly affect the tendency of the call option prices.

\begin{figure}[h!]
\begin{center}
\includegraphics[height=2in, width=6in]{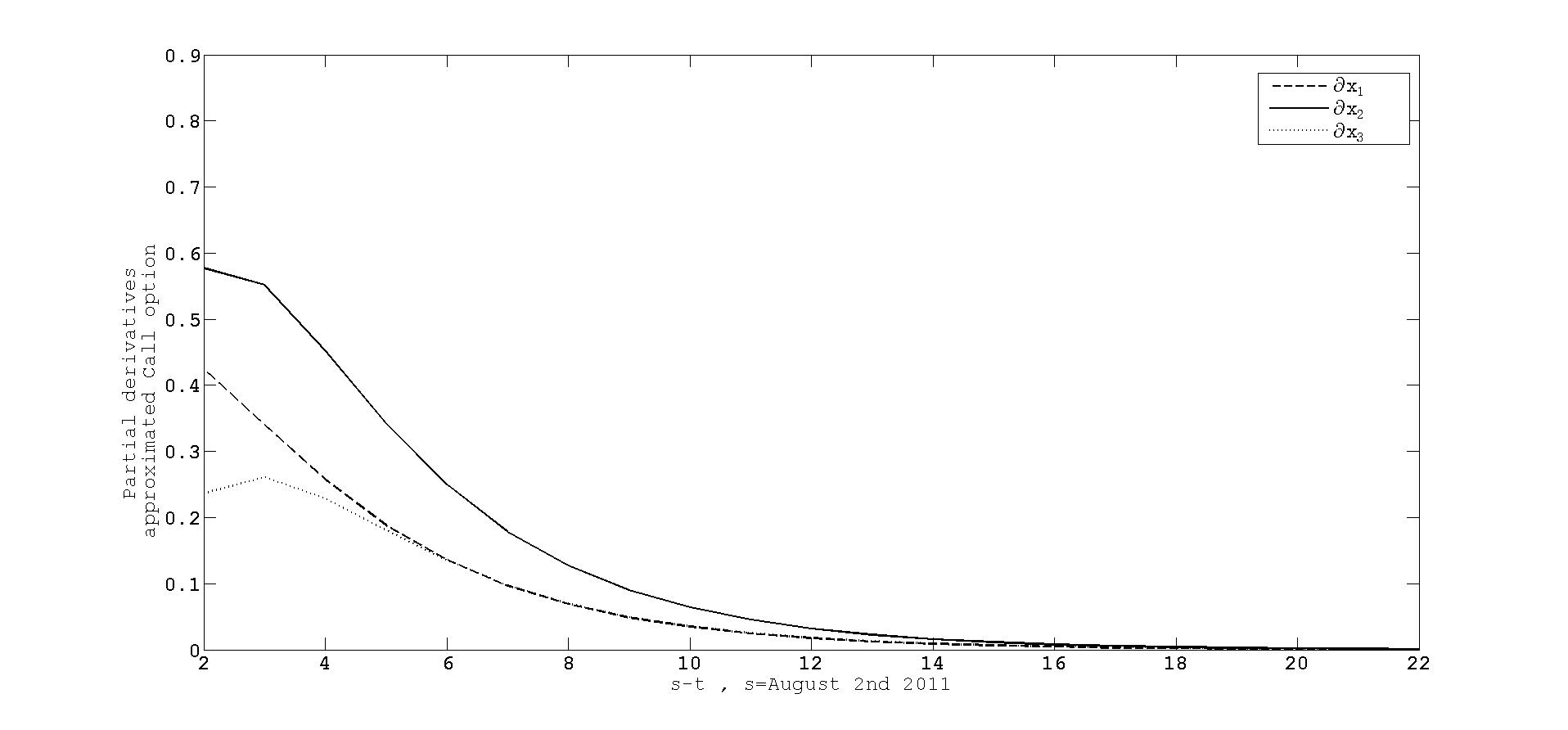}
\caption{$(\partial\widetilde{C}(t,\tau,s,\widetilde{F}_{\text{CDD}}(t,s,\texttt{x}_1,\ldots,\texttt{x}_p))/\partial \texttt{x}_{i})_{\vert_{\mathbf{x}=\mathbf{0}}}$
for $i=1,2,3$ with exercise time $\tau$ being August 1st, 2011 and measurement day $s$ August 2nd, 2011.}
\label{fig:dappC000}
\end{center}
\end{figure}

Figure.~\ref{fig:dappC000} shows the partial derivatives of the approximate call option prices
with respect to the coordinates of $\mathbf{x}$ derived in Proposition. \ref{pro:dappC-ts} when $\mathbf{X}(t)=\mathbf{0}$.
We observe that the partial derivatives in Figure.~\ref{fig:dC000} and Figure.~\ref{fig:dappC000} show a close habavior. 
We also see that for small times to maturity the partial derivatives of the approximate call option price are bigger than 
the partial derivatives of the call option price. 

We end our analysis with the results for the case $\mathbf{X}(t)=\mathbf{e}_1$.

\begin{figure}[h!]
\begin{center}
\includegraphics[height=2in, width=6in]{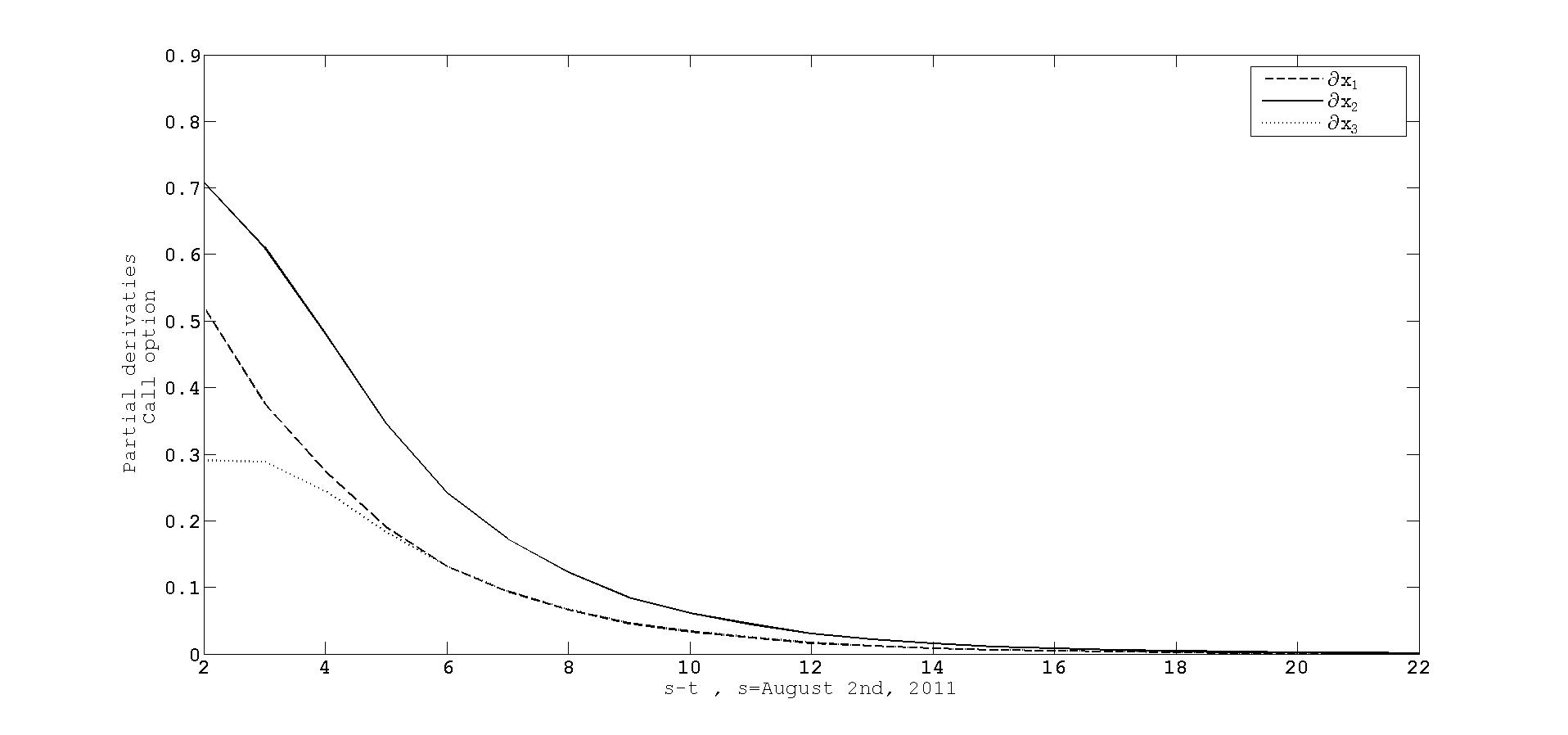}
\caption{$(\partial C(t,\tau,s,\texttt{x}_1,\ldots,\texttt{x}_p))/\partial\texttt{x}_i)_{\vert_{\mathbf{x}=\mathbf{e}_{1}}}$
for $i=1,2,3$ with exercise time $\tau$ being August 1st, 2011 and measurement day $s$ August 2nd, 2011.}
\label{fig:dC100}
\end{center}
\end{figure}

\begin{figure}[h!]
\begin{center}
\includegraphics[height=2in, width=6in]{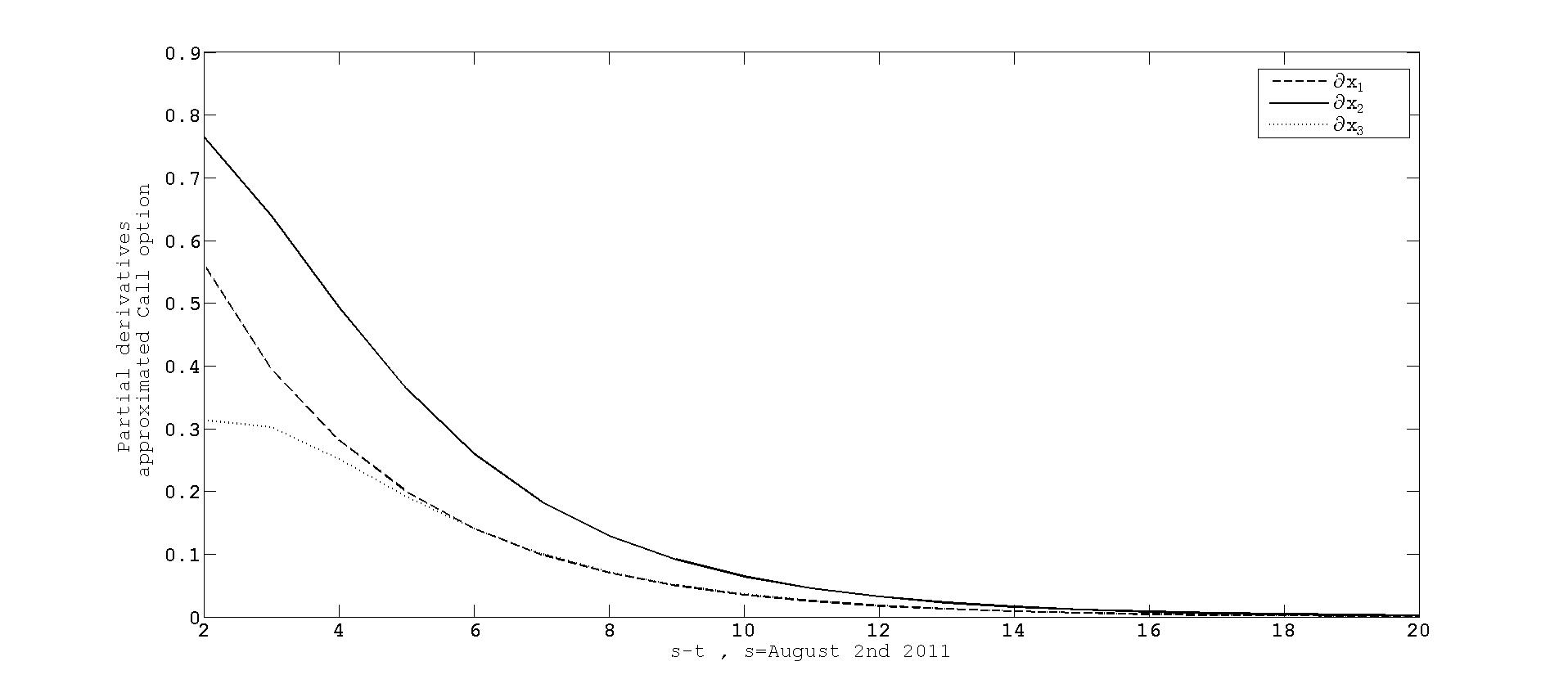}
\caption{$(\partial\widetilde{C}(t,\tau,s,\widetilde{F}_{\text{CDD}}(t,s,\texttt{x}_1,\ldots,\texttt{x}_p))/\partial \texttt{x}_{i})_{\vert_{\mathbf{x}=\mathbf{e_1}}}$
for $i=1,2,3$ with exercise time $\tau$ being August 1st, 2011 and measurement day $s$ August 2nd, 2011.}
\label{fig:dappC100}
\end{center}
\end{figure}

Figure. \ref{fig:dC100} and Figure.\ref{fig:dappC100} show the partial derivatives for the (approximate) call option prices, respectively. 
We see also here that both prices are more sensitive to changes in the second component of $\mathbf{x}$ in all the domain. Furthermore,
the sensitivity to this component decreases as time to maturity increases.

\section{Sensitivity of CDD futures prices with measurement over a  period}\label{sec:appendix}
The sensitivity analysis of CDD future prices with measurement over a period $[\tau_1,\tau_2]$ with respect to infinitesimal changes in the components of $\mathbf{X}(t)$ 
can be performed by means of the partial derivatives provided in Proposition.\ref{prop:dFCDD-ttau1tau2} and Proposition. \ref{prop:dappFCDD-ttau1tau2}.
We proceed analogously as in the previous section for CDD futures prices with a measurement day.

The random variable $\Phi(Z(t,s))$ with $Z(t,s)$ in \eqref{Z(t,s)} makes here also the difference between the results provided in both Propositions. The same reasoning followed for CDD future prices with a measurement day is valid to conclude that 
\begin{equation}
\label{relation1}
\big(\frac{\partial}{\partial\texttt{x}_i}F_{\text{CDD}}(t,\tau_1,\tau_2,\texttt{x}_1,\texttt{x}_2,\texttt{x}_3)\big)_{\vert_{\mathbf{x}=\mathbf{X}(t)}}
\leq
\frac{\partial}{\partial\texttt{x}_i}\widetilde{F}_{\text CDD}(t,\tau_1,\tau_2,\texttt{x}_1,\texttt{x}_2,\texttt{x}_3).
\end{equation}
Hence, the approximate CDD futures price becomes more sensitive to any infinitesimal change in the coordinates of $\mathbf{X}(t)$ than the CDD futures price.

\begin{figure} [h!]
\includegraphics[height=2in, width=6in]{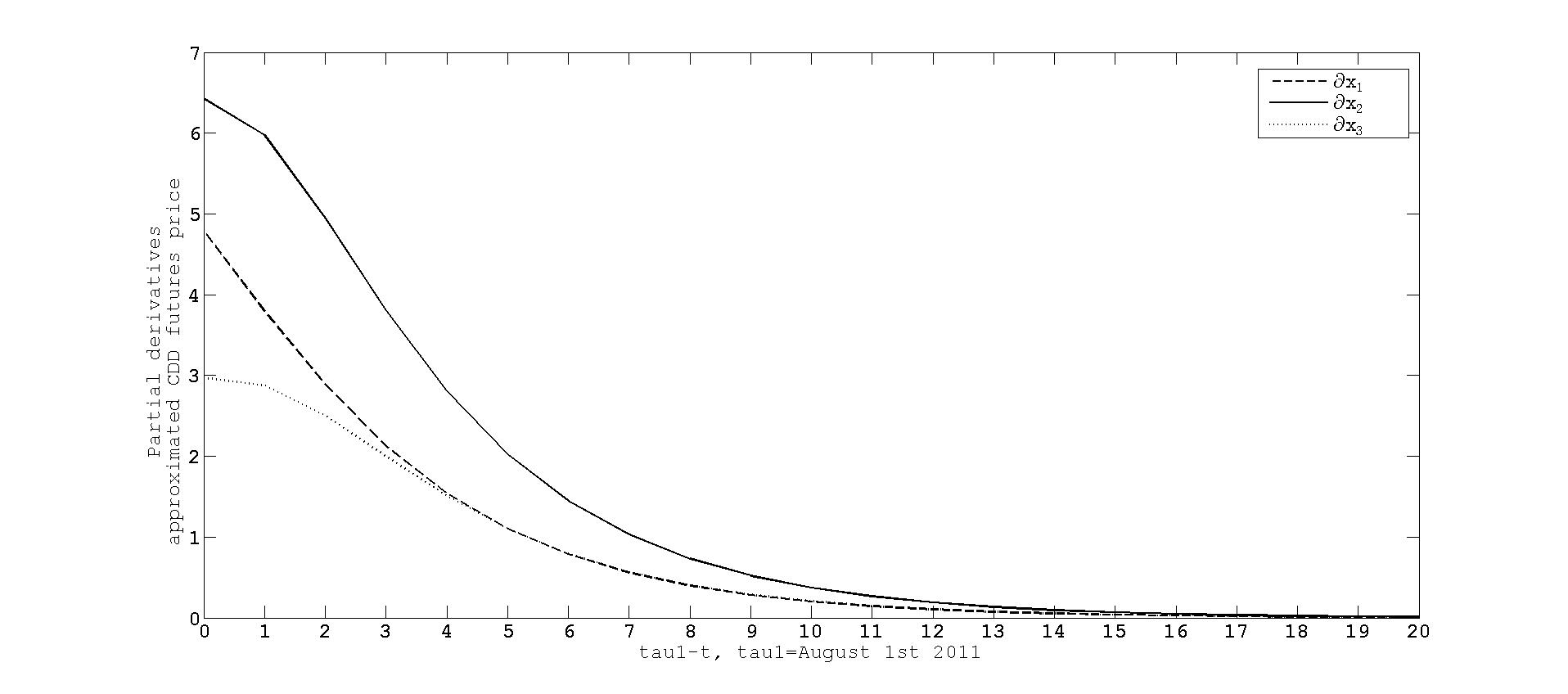}
\caption{$\partial\widetilde{F}_{\text{CDD}}(t,\tau_1,\tau_2,\texttt{x}_1,\ldots,\texttt{x}_p)/\partial\texttt{x}_i$ 
as a function of $\tau_1-t$ with measurement period $[\tau_1,\tau_2]$ being August, 2011.}
\label{fig:dappFCDD-mp}
\end{figure}

Figure.\ref{fig:dappFCDD-mp} shows the partial derivatives of the approximate CDD futures price with respect the coordinates of  $\mathbf{x}$ derived in 
Proposition.~\ref{prop:dappFCDD-ttau1tau2}.

The x-axis considers the time to maturity, $\tau_1-t$, where $\tau_1$ is August 1st, 2011
and the measurement period $[\tau_1,\tau_2]$ is August 2011. 
The y-axis shows the different partial derivatives of the approximate CDD futures price. 
Firstly, we observe that the values here of the partial derivatives 
are greater than those corresponding to the partial derivatives of the approximate CDD futures price with a measurement day,
with more emphasis on times $t$ which are close to $\tau_1$. 
Indeed, the partial derivatives of the approximate CDD futures prices can be understood 
as the partial derivatives of the approximate CDD futures prices with measurement day $s$ which runs over $[\tau_1,\tau_2]$ as shown as follows
$$
\frac{\partial}{\partial\texttt{x}_i}\widetilde{F}_{\text{CDD}}(t,\tau_1,\tau_2,\texttt{x}_1,\ldots,\texttt{x}_p)\nonumber
=\int_{\tau_1}^{\tau_2}\frac{\partial}{\partial\texttt{x}_i}\widetilde{F}_{\text{CDD}}(t,s,\texttt{x}_1,\ldots,\texttt{x}_p)\,ds\,.
$$

Recall that Figure.~\ref{fig:dappFCDD} shows that the derivatives of the approximate CDD futures price with a measurement day are 
positive and when time to maturity increases tend to zero. This fact together with the 
relation in \eqref{relation1} let us to justify why the derivatives of the approximate CDD futures prices with measurement over a period behave in this way.
We also observe that any infinitesimal change in $\texttt{x}_2$ dominates more the behaviour of the  
approximate CDD futures price, followed by any change in $\texttt{x}_1$ and $\texttt{x}_3$. 
This was exactly the same tendency followed by the partial derivatives of the approximate CDD futures price with
a measurement day, see Figure.~\ref{fig:dappFCDD}, for times to maturity greater than 2.  

For the study of the sensitivity of the partial derivatives of the CDD futures price with measurement over a period 
derived in Proposition. \ref{prop:dFCDD-ttau1tau2} we consider the cases $\mathbf{X}(t)=\mathbf{0}$ and $\mathbf{X}(t)=\mathbf{e}_1$.
 
\begin{figure}[h!]
\begin{center}
\includegraphics[height=2in, width=6in]{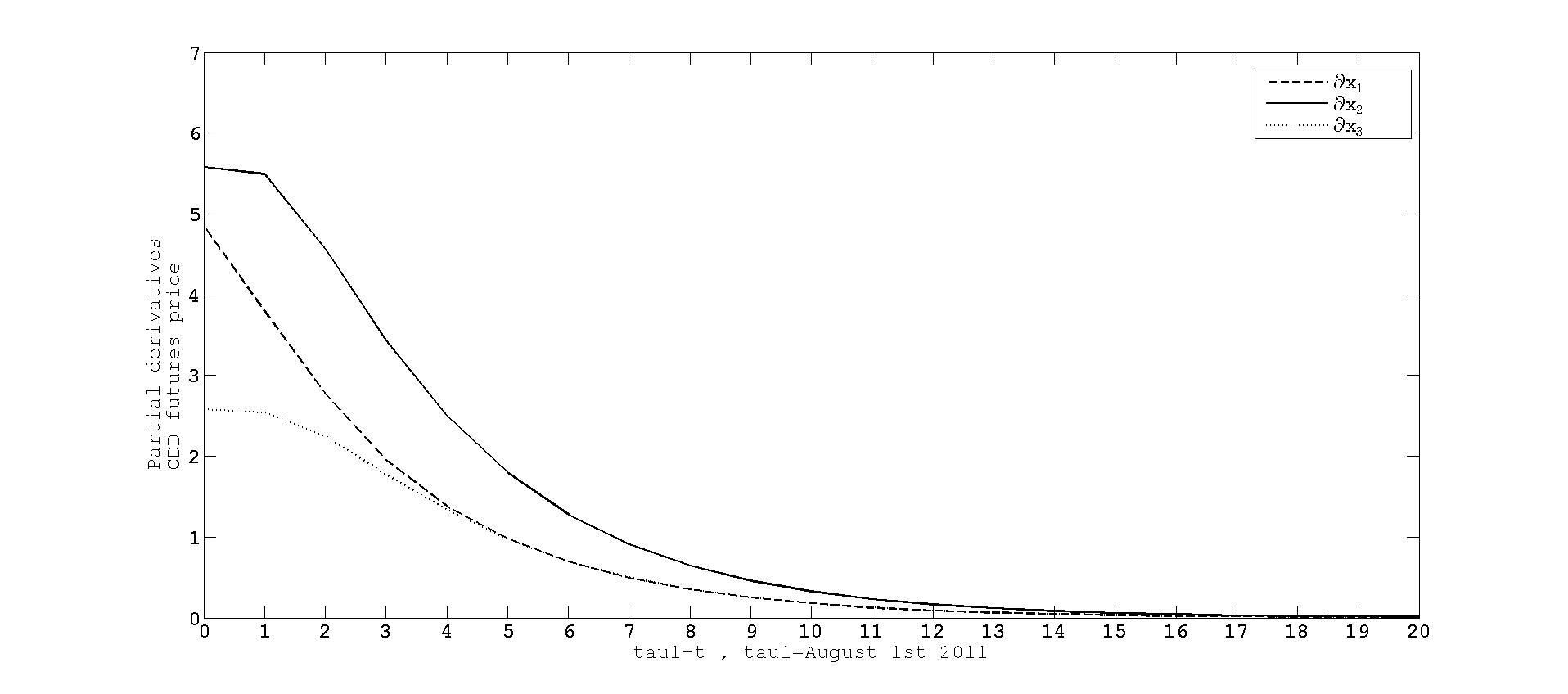}
\caption{$(\partial F_{\text{CDD}}(t,\tau_1,\tau_2,\texttt{x}_1,\ldots,\texttt{x}_p)/\partial\texttt{x}_i)_{\vert_{\mathbf{x}=\mathbf{0}}}$ for $i=1,2,3$
as a function of $\tau_1-t$ with measurement over a period $[\tau_1,\tau_2]$ being August, 2011.}
\label{fig:dFCDD000-mp}
\end{center}
\end{figure}

\begin{figure}[h!]
\begin{center}
\includegraphics[height=2in, width=6in]{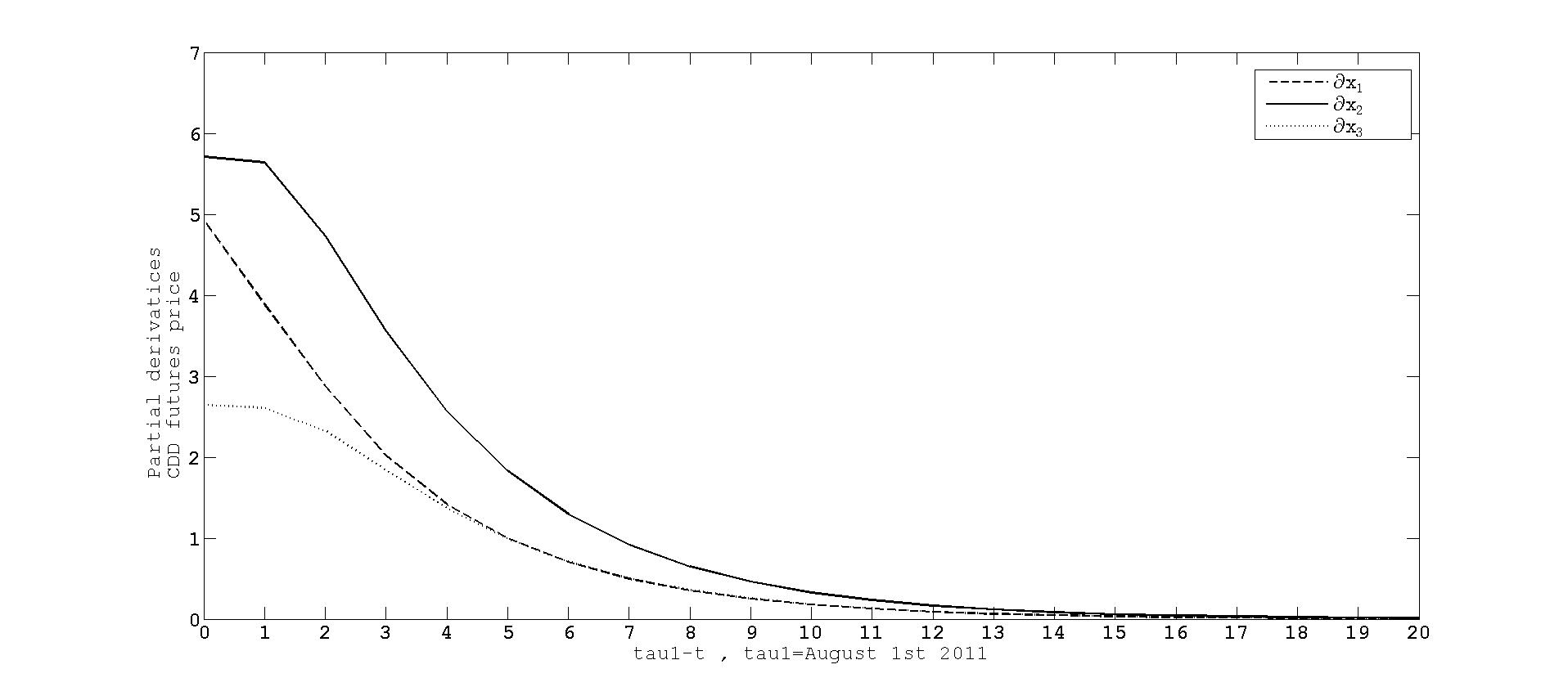}
\caption{$(\partial F_{\text{CDD}}(t,\tau_1,\tau_2,\texttt{x}_1,\ldots,\texttt{x}_p)/\partial\texttt{x}_i)_{\vert_{\mathbf{x}=\mathbf{e}_{1}}}$ for $i=1,2,3$
as a function of $\tau_1-t$ with measurement over a period $[\tau_1,\tau_2]$ being August, 2011.}
\label{fig:dFCDD100-mp}
\end{center}
\end{figure}

Figure. \ref{fig:dFCDD000-mp} and Figure. \ref{fig:dFCDD100-mp} show also that perturbations in $\texttt{x}_2$
affect also more in these cases the CDD futures price.

\section{Conclusions}
In this paper we have studied the local sensitivity of the (approximate) CDD and HDD futures and options prices
with respect to a perturbation in the deseasonalized temperature or in one of its derivatives up to a certain order,
determined by the CAR process modelling the deseasonalized temperature. We have considered the partial derivatives
of these financial contracts with respect to these variables (deseasonalized temperature and its derivatives). 
The HDD and CDD futures and call option prices and their approximate formulas were derived in Benth and Solanilla Blanco \cite{BenSol-A3}.
We have considered and empirical analysis where we have taken 
the same CAR(3)-process fitted to the time series of New York temperatures in Benth and Solanilla Blanco \cite{BenSol-A3}.
The sensitivity study of these financial contracts with a fixed measurement day shows first that the approximate futures prices 
are more sensitive to any perturbation in one of these variables than the theoretical futures prices. Nevertheless, the relative error 
between both partial derivatives is rather small.
We also observe that one time prior to the considered measurement day the behaviour of the (approximate) futures prices is more affected by 
a perturbation in the deseasonalized temperature. As time to maturity increases then a perturbation in the slope of the deseasonalized temperature
dominates the behaviour of both futures prices. At the long end any perturbation of these variables hardly affect
their behaviour. For the call option prices we also observe that the approximate model is more sensitive to any pertubation in one of the previous variables than the theoretical model. We emphasize that unlike futures prices,
any perturbation in the slope of the deseasonalized temperature, dominates the behaviour of the call option prices.
We have also extended the analysis of sensitivity to futures prices with measurement over a fixed month. We have seen that in this case
the slope of the deseasonalized temperature dominates the bahaviour in all the domain.


\bibliography{A4_Arxiv_v1}
\bibliographystyle{plain}


\end{document}